\documentclass[]{aastex63}

\usepackage{graphicx}
\usepackage{amsmath}
\usepackage{bm}
\usepackage{amssymb}
\usepackage{romannum}
\usepackage{textcomp}
\usepackage{upgreek}
\submitjournal{PSJ}
\shorttitle{Geometric Albedo and Phase Function of Earth}
\shortauthors{Robinson}
\graphicspath{{./}{figures/}}
\begin{document}

%%%
%
\pagenumbering{arabic}
%
%%%

%%%
\title{Inferring and Interpreting the Visual Geometric Albedo and Phase Function of Earth}
%%%

%%%
\correspondingauthor{Tyler D. Robinson}
\email{tdrobin@arizona.edu}
%%%

%%%
%
\author[0000-0002-3196-414X]{Tyler D. Robinson}
\affiliation{Lunar \& Planetary Laboratory, University of Arizona, Tucson, AZ 85721 USA}
\affiliation{Habitability, Atmospheres, and Biosignatures Laboratory, University of Arizona, Tucson, AZ 85721, USA}
\affiliation{NASA Nexus for Exoplanet System Science Virtual Planetary Laboratory, University of Washington, Box 351580, Seattle, WA 98195, USA}
%
%%%

%%%
% AAS Journals Limit: 250 words
\begin{abstract}
Understanding reflectance-related quantities for worlds enables effective comparative planetology and strengthens mission planning and execution. Measurements of these properties for Earth, especially its geometric albedo and phase function, have been difficult to achieve due to our Terrestrial situation\,---\,it is challenging to obtain planetary-scale brightness measurements for the world we stand on. Using a curated dataset of visual \added{(0.4--0.7\,$\upmu$m)} phase-dependent, disk-averaged observations of Earth taken from the ground and spacecraft, alongside a physical-statistical model, this work arrives at a definitive value for the visual geometric albedo of our planet: $0.242^{+0.005}_{-0.004}$. This albedo constraint is up 30--40\% smaller than earlier, widely-quoted values. The physical-statistical model enables retrieval-like inferences to be performed on phase curves, and includes contributions from optically thick clouds, optically thin aerosols, Rayleigh scattering, ocean glint, gas absorption, and Lambertian surface reflectance. Detailed application of this inverse model to Earth's phase curve quantifies contributions of these different processes to the phase-dependent brightness of the Pale Blue Dot. Model selection identifies a scenario where aerosol forward scattering results in a false negative for surface habitability detection, \added{which implies that aerosol forward scattering can effectively mimic an ocean glint signature in broadband visual phase curves.} Observations of phase curves for Earth at redder-optical or near-infrared wavelengths could disentangle ocean glint effects from aerosol forward scattering. \added{Finally, a review of albedos and planetary photometry is provided as well as a simple two-parameter fit to Earth's visual phase curve to ease adoption into other tools.}
\end{abstract}
%
%%%

%%%
% to do:
% [] 
%%%

%%%
%
%\keywords{Remote sensing (2191) --- Radiative transfer simulations (1967) --- Exoplanets (498) --- Earth (planet) (439)}
%
%%%

%%%
%
\section{Introduction} \label{sec:intro}
%
%%%

A core science goal for NASA's upcoming Habitable Worlds Observatory \citep[HWO;][]{feinbergetal2024} is the detection and characterization of roughly 25 potentially Earth-like worlds orbiting nearby stars, as recommended by the National Academies' 2020 Decadal Survey on Astronomy and Astrophysics \citep{decadal2020}. The brightness of the HWO exo-Earth targets as a function of phase angle (i.e., star-planet-observer angle) is of central importance to both the survey and characterization goals of the mission \citep{morganetal2022,starketal2024}. A reflected-light survey for exo-Earths depends on how bright these targets are at different phase angles while photometry and spectroscopy at a variety of phases would enable better atmospheric characterization \citep{nayaketal2017}.

The phase-dependent brightness of a world is a long-studied topic in planetary science \citep[e.g.,][]{knucklesetal1961}. Typically brightness as a function of phase angle is represented through two separate quantities: the geometric albedo and the planetary phase function. The geometric albedo, which is detailed more rigorously below, measures the brightness of a world at full phase (i.e., a phase angle of 0\textdegree). The phase function then translates this brightness to other phase angles and is normalized to unity at full phase. Both the geometric albedo and the planetary phase function can be wavelength-dependent quantities, or both can be wavelength-averaged over some bandpass.

Analyses of spacecraft-measured phase functions yielded early constraints on the particle sizes and vertical distribution of aerosols for outer solar system worlds \citep{tomaskoetal1980b,sromovskyetal1981,tomasko&smith1982,pollacketal1986}. More recently, sulfuric acid cloud properties for Venus have been constrained by phase curve observations \citep{mallamaetal2006,garciamunozetal2014}, where phase curves in polarized light measured from the ground had originally been used to infer the chemical properties of Cytherean clouds \citep{hansen&hovenier1974}. \added{Polarimetric techniques applied to Earthshine observations have also been shown to constrain properties of Earth's water clouds \citep{sterziketal2020}.} Disk-integrated photometry of Titan from NASA's \textit{Cassini} mission revealed the impressive fact that, due to haze forward scattering, Titan can be brighter at extreme crescent phases than at full phase \citep{garciamunozetal2017,cooperetal2025}. Additionally, \textit{Cassini} observations of Jupiter revealed significant non-Lambertian scattering effects in the derived phase curves \citep{mayorgaetal2016,heng&li2021,jonesetal2025}.

Despite (or, more realistically, because of) Earth being our home planet, robust measures of Earth's geometric albedo and phase function remain elusive. The latter has been best revealed through studies of Earth-light reflected by the observable portion of the Moon not illuminated by the Sun \citep[``la Lumi\'ere cendr\'ee,'' or ashen light, to early scientists;][]{danjon1928}, now commonly called Earthshine \citep{goodeetal2001}. Often, though, Earthshine studies aim to integrate their observations over phase angle to infer a spherical albedo \citep{qiuetal2003,palleetal2003,palleetal2004,goodeetal2021}, which is a key quantity for understanding Earth's radiative energy budget. Thus, the phase function is not generally reported or highlighted.

Estimates for Earth's geometric albedo have also stemmed from broadband visual Earthshine observations. These estimates vary widely and are generally not reported with uncertainties or constraints on variability. A value of 0.367 is stated in \textit{Allen's Astrophysical Quantities} \citep{cox2000}, stemming from century-old visual Earthshine data \citep{danjon1928,danjon1954,harris1961}. In work focused on photometry of Mercury, Venus, and Mars, \citet{mallama2009} extrapolated broadband Earthshine data to full phase and estimate a geometric albedo of 0.2. In subsequent efforts this value was updated to 0.434 \citep[in V-band;][]{mallamaetal2017}.

While additional constraints on Earth's geometric albedo and phase function are plainly needed, so too are new methods for statistically analyzing reflected-light phase curves for constraints on planetary environmental characteristics. Valuable information for solar system planetary environments from the works cited above benefit from applying detailed (and often sophisticated) reflectance models that are tailored to the world at hand. Such luxuries are unlikely to be an option for directly-imaged HWO targets. Efforts below build upon techniques from \citet{heng&li2021} and \citet{jonesetal2025}, where wavelength-dependent scattering properties are inferred from independent treatments of spectral phase curves of Jupiter (and Enceladus). A novel development in this manuscript is to connect wavelength dependence to different physical properties (e.g., cloudiness, ocean coverage) in spatially-resolved models so that reflected-light phase curves can then be used to elucidate the planetary environment\,---\,an approach that is increasingly common in the analysis of hot Jupiter emitted-light phase curves where rotational locking connects the phase angle to the observable range of planetary longitudes \citep{fengetal2016,tayloretal2020,macdonaldetal2020,chubb&min2022}.

Reaching clarity on Earth's geometric albedo and phase-dependent brightness is important and timely, especially given the central role Earth\,---\,as the quintessential Earth-like world\,---\,will play in the development of HWO. The work below aims to provide this clarity, in addition to developing new analysis techniques. As relevant quantities that measure ``brightness,'' ``reflectivity,'' or ``albedo'' are often poorly defined, Section~\ref{sec:albedos} and Appendix~\ref{sec:photom_review} review and develop the theory required to formally define key quantities. Observational data at broadband visual wavelengths are curated in Section~\ref{sec:data} and used to constrain physical-statistical models described in Section~\ref{sec:model}. The constrained models are then used to obtain statistical inferences of fundamental planetary properties of Earth in Section~\ref{sec:results} and that are discussed in Section~\ref{sec:discuss}.

%%%
%
\section{The A-Team: A Summary of Albedos} \label{sec:albedos}
%
%%%

A variety of quantities related to reflectivity are used in planetary exploration, and occasionally these terms receive the broad label of ``albedo.'' Before presenting reflectivity data for Earth, then, it is useful to define and summarize the key reflectance- and albedo-related quantities. Appendix~\ref{sec:photom_review} contains a more detailed review that begins with surface reflectance properties and building up to planetary disk-averaged quantities. An earlier explanation of planetary photometry is provided by \citet{lesteretal1979} and a thorough exploration of analytic models for the phase-dependent brightness of planetary targets is presented in \citet{madhusudhan&burrows2012}.

Quantities that are often simply labeled as ``albedo'' include: a surface flux albedo ($A$), the geometric albedo ($A_{\rm g}$), the spherical albedo ($A_{\rm s}$), the Bond albedo ($A_{\rm B}$), and the apparent albedo ($A_{\rm app}$). Excepting the Bond albedo, these quantities can, in general, be wavelength-dependent. The surface flux albedo describes how effectively a surface location on a world reflects incident flux, and can depend on whether the incident flux is collimated or diffuse. The remaining albedos are most relevant to the work that follows as they are measured for entire planetary disks or globes.

The geometric albedo is a measure of the full-phase reflectivity of a planet, which is inferred by measuring (or modeling) the average intensity of reflected light over the entire planetary disk ($\bar{I}$) at a phase angle of zero and dividing this by the solar (or stellar) flux ($F_{\rm s}$) incident normally at the planetary orbital distance ($r$), as given in Equation~\ref{eqn:geomalb}. The geometric albedo can be translated into a disk-averaged intensity at any other phase angle ($\alpha$)  using the planetary phase function, $\Phi \left( \alpha \right)$, as indicated in Equation~\ref{eqn:phasefunc}. This phase-dependent brightness is useful in exoplanet science as it appears in the canonical expression for the planet-to-star flux ratio,
\begin{equation}
    \frac{F_{\rm p}}{F_{\rm s}} = A_{\rm g} \Phi\left( \alpha \right) \left( \frac{R_{\rm p}}{r} \right)^2 \ ,
\end{equation}
where $R_{\rm p}$ is the planetary radius.

The spherical albedo is a wavelength-dependent quantity relevant to planetary radiative balance that is the ratio of the hemispherically-integrated power reflected by a planet to the power intercepted by the planet. As indicated in Equation~\ref{eqn:sphere_alb}, the upwelling reflected-light flux emergent from all locations on a planet must be known to infer a spherical albedo. If the spherical albedo is weighted by the incident solar (or stellar) spectral energy distribution and integrated over all wavelengths, one obtains a quantity related the Bond albedo (Equation~\ref{eqn:bond_alb}), which is a bolometric quantity relevant to the overall planetary radiative balance.

Finally, the apparent albedo is a useful quantity for describing the phase-dependent reflectivity of a planet. Formally defined in Equation~\ref{eqn:Aapp}, this albedo is obtained by scaling the planetary disk-averaged intensity by the intensity that would be recorded for a Lambert sphere with the same illumination geometry. Put another way, the apparent albedo is the surface flux albedo required for a Lambert sphere to reproduce the disk-averaged intensity observed for a planet at a given phase angle. The utility of the apparent albedo\,---\,as is highlighted throughout the Earth analyses below\,---\,is that it scales out uninteresting first-order illumination effects and, thus, better highlights non-isotropic scattering behaviors from planetary disks especially at high phase angles. \added{At full phase, the apparent albedo is related to the geometric albedo through Equation~\ref{eqn:Aapp}, with,}
\begin{equation}
    A_{\rm g} = \frac{2}{3} A_{\rm app} \left(\alpha = 0 \right) \ .
\end{equation}

%%%
%
\section{Data} \label{sec:data}
%
%%%

Only a limited number of different missions and techniques have provided measurements of the phase-dependent, disk-averaged brightness of Earth \citep{robinson&reinhard2020}. To date, only Earthshine observations have provided broad coverage in phase angle. As the Earthshine measurements are typically broadband visual (i.e., 0.4--0.7\,$\upmu$m), all other data sources discussed below require an integration across wavelength to best compare to Earthshine data. Similarly, as Earthshine observations are generally insensitive to rotational variability (owing to these observations being taken once per night in an intermittent fashion), time-resolved data sources are averaged to remove sensitivity to rotational effects and, thus, better compare with the Earthshine data.

\begin{figure}
    \centering
    \includegraphics[scale=0.75,trim=0mm 0mm 0mm 0mm]{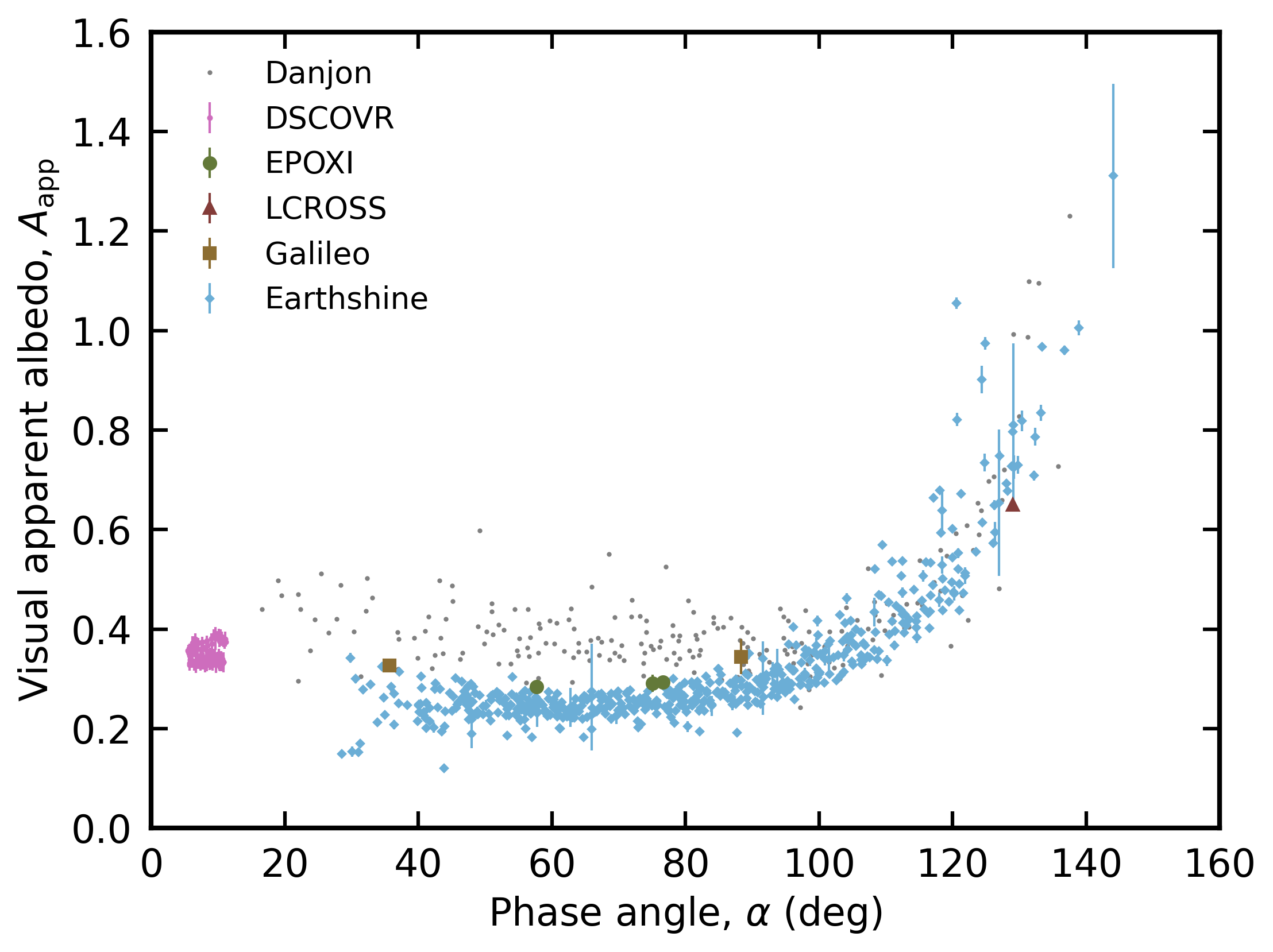}
    \caption{Phase-dependent measurements of Earth's broadband visual \added{(0.4--0.7\,$\upmu$m)} apparent albedo, including historic observations from \citet{danjon1928}. Mission or observing technique is indicated by datapoint color and shape. Uncertainties are indicated, which are sometimes smaller than the point size. \added{As a reminder, the geometric albedo is equal to 2/3 of the apparent albedo at full phase.}}
    \label{fig:earth_data}
\end{figure}

Subsections below detail data sources and any treatments applied to these data. Figure~\ref{fig:earth_data} shows the final curated dataset and includes, for completeness, historical observations from \citet{danjon1928}. Now-understood biases in the Danjon data prevent their inclusion in the subsequent analyses \citep{qiuetal2003,palleetal2003}. %Note that the curated data are shown as apparent albedo; this helps to better highlight structure at high phase angles and this measure will be used for most subsequent plots of phase-dependent data and models.
In the material that follows, relative versus absolute changes in albedo quantities will be distinguished for clarity (which is important as some authors report albedo-related quantities as percentages). For example, when comparing albedo measurements of 0.20 and 0.24, the latter has an albedo increase of 0.04 over the former, which represents a 20\% increase in reflectivity.

\subsection{Earthshine}
A total of 531 phase-dependent, broadband visual Earthshine measurements were provided by earlier works \citep{qiuetal2003,palleetal2003}. An inspection of this data collection reveals a small number of points with unphysically small apparent albedos. To remove these outliers, each datapoint was assigned a standard score based on its 20 nearest neighbors in phase angle and all points with a standard score above 4 were removed. This culls seven points (i.e., 1.3\% of the overall dataset), most of which have an apparent albedo of 0.01 or smaller. The original data are reported with photometric errors, and systematic errors are not incorporated into any following analyses as comparisons between Earthshine observations and data from Earth-observing satellites have shown the former to be accurate to about 1\% \citep[relative;][]{palleetal2016}.

\subsection{DSCOVR}
The NASA \textit{Deep Space Climate Observatory} (\textit{DSCOVR}) mission observes Earth at near-full phase from the first Earth-Sun Lagrange point and includes narrowband, spatially-resolved imaging of Earth from the Polychromatic Imaging Camera. Data adopted for the analyses below come from earlier studies of Earth as an analog for an exoplanet \citep{jiangetal2018,fanetal2019,aizawaetal2020,guetal2021} wherein the spatially-resolved EPIC images were integrated over the disk of Earth to produce disk-averaged brightness measurements. As the Earthshine observations are, on average, spaced 4 days apart (and are, thus, generally insensitive to rotational variability), the roughly 5,000 EPIC observations (per bandpass) were averaged over four-day intervals to remove rotational effects. The time-averaging process resulted in 94 phase-dependent measurements of disk-averaged brightness in each EPIC filter. To mimic an Earthshine observation\,---\,that, in this setup, would obtain only a single snapshot at some point over the four-day period\,---\,the rolling root-mean-square spread in the EPIC photometry is used to assign an ``uncertainty'' to each bandpass-dependent datapoint. Thus, a \textit{DSCOVR} datapoint statistically represents a single observation that could have occurred at any point over the four-day averaging window.

The EPIC filters at 442\,nm, 551\,nm, 680\,nm, and 688\,nm (with bandwidths of 3\,nm, 3\,nm, 2\,nm, and 1\,nm, respectively) span the visual range and disk-averaged intensity values in these four filters were combined via an intensity-weighted average to produce approximate broadband visual disk-averaged brightness measurements. A suite of 72 high-fidelity, high-spectral resolution simulations of the disk-averaged brightness of Earth \citep{robinsonetal2011} were selected from a larger dataset \citep{robinsonetal2010} to capture the spread of Earth phase angles seen by \textit{DSCOVR}. Integrating these high-fidelity models across the full visual range and comparing this broadband brightness measurement to a case where the weighted sum across the EPIC filter treatment was applied reveals that the latter produces a systematic overestimation of brightness of about 1\% (relative). No additional treatments for this potential biasing was considered as it is both small and comparable to the calibration accuracy of the instrument \citep{haneyetal2022}. As in the Earthshine case, systematic errors for the EPIC data are not considered in the analyses that follows due to their generally small size.

\added{The \textit{DSCOVR} observations provide an opportunity to efficiently estimate the visual geometric albedo of Earth that is anticipated from the analysis that follows below, as the EPIC observations are all near full phase (where the geometric albedo is defined). Adjusting the \textit{DSCOVR} observations to full phase using the Lambert phase function (which is typically a 1--2\% adjustment) yields a visual geometric albedo of 0.23, averaged over the entire dataset. This estimate is already markedly smaller than most values quoted above from earlier studies.}

\subsection{EPOXI}
The repurposed NASA \textit{Deep Impact} flyby spacecraft obtained spatially-, temporally-, and spectrally-resolved observations of the distant Earth on five separate occasions, as part of the \textit{EPOXI} mission \citep{livengoodetal2011}. Three of these datasets were at equator-on viewing geometries while two were pole-on. The latter do not compare well to the more equator-on Earthshine and \textit{DSCOVR} observations and are markedly more-reflective. Thus, the pole-on pointings were omitted from the subsequent analyses.

Optical photometry from the three equator-on pointings (phase angles of 57.7\textdegree, 75.1\textdegree, and 76.6\textdegree) were rotationally-averaged and combined via a flux-weighted average to produce visual disk-averaged brightnesses. Like with the \textit{DSCOVR} data, the root-mean-square variability over a rotation is used to indicate a brightness measurement ``uncertainty'' analogous to an Earthshine observation. The filters spanning the visual range for the High-Resolution Instrument camera for \textit{Deep Impact} were absolutely calibrated to 5\% \citep{klaasenetal2008}, so a systematic offset factor for the \textit{EPOXI} datapoints is included as a rather uninteresting fitted parameter in the analyses below (at least as compared to any fitted physical parameters). \added{Such calibration offsets are a viable explanation for why some spacecraft observations in Figure~\ref{fig:earth_data} appear systematically offset from the well-calibrated Earthshine data.}

\subsection{Galileo}
Optical imaging of Earth by NASA's \textit{Galileo} spacecraft was acquired during gravitational assists in 1990 and 1992 (at phase angles of 35.7\textdegree~and 88.3\textdegree, respectively). Rotation- and disk-averaged photometry from these assists are provided by \citet{straussetal2024}. Similar to the \textit{EPOXI} treatment, a flux-weighted average is applied to the filter photometry to generate measurements in the visual band and the analyses below include fitting for a potential systematic calibration bias of about 8\% \citep{klaasenetal1999}.

\subsection{LCROSS}
Observations from NASA's \textit{Lunar CRater Observation and Sensing Satellite} (\textit{LCROSS}) included pointings to Earth to acquire spectroscopy at ultraviolet and visible wavelengths (0.26--0.65\,$\upmu$m) at high resolving power. The field-of-view of the spectrometer was smaller than the apparent size of Earth's disk, which was especially problematic for gibbous phase pointings. A crescent phase pointing, though, captured most of the illuminated disk, save for the low-intensity crescent horns. A high fidelity, bespoke model of this crescent phase observation \citep{robinsonetal2014lcross} was used to correct the single datapoint for the missing crescent horns (a 12\% enhancement) and to extend the spectral observation to include the 0.65--0.70\,$\upmu$m range (a 16\% enhancement). As with the \textit{EPOXI} and \textit{Galileo} data, subsequent analyses include a fitted systematic offset for the \textit{LCROSS} datapoint, whose absolute calibration was accurate to about 10\% \citep{robinsonetal2014lcross}.

%%%
%
\section{Models} \label{sec:model}
%
%%%

The time- and phase-dependent Earth visual \added{(0.4--0.7\,$\upmu$m)} apparent albedo data were fit with a physical-statistical model that captures the physics of relevant atmospheric and surface scattering processes while representing variability as a statistical process. At a given phase angle, the probability density for apparent albedo, $p(A_{\rm app})$, is represented as,
\begin{equation}
    p\left( A_{\rm app} \right) = p_{\rm G}\left(A_{\rm app} | A_{\rm m}(\alpha;\bm{x}), \sigma_{\rm m}(\alpha;\bm{x}) \right) \ ,
    \label{eqn:physstatmdl}
\end{equation}
where $p_{\rm G}$ is the Gaussian (or normal) distribution, $A_{\rm m}$ is the physical model for the apparent albedo, $\sigma_{\rm m}$ is the model for the standard deviation due to variability, and $\bm{x}$ is the vector of fitted parameters. Model components, the treatment of data systematics, adopted priors, and the likelihood function are detailed below.

\subsection{Physical Apparent Albedo Model}
Phase- and wavelength-dependent treatments of Lambertian surface scattering, ocean glint, Rayleigh scattering, optically thick cloud scattering, vertically optically thin aerosol scattering, and ozone absorption are incorporated into a physical forward model. A spectral treatment is warranted as the Rayleigh scattering optical depth increases monotonically by an order of magnitude across the 0.4--0.7\,$\upmu$m range while ozone opacity varies by a factor of more than 1,000 across this same range, peaking at $0.58$\,$\upmu$m. Both molecular oxygen and water vapor have very weak and narrow features in this wavelength range for Earth \citep{robinsonetal2014lcross}, but are omitted from the subsequent analyses owing to ozone being the dominating gaseous absorber in this spectral range. The physical treatments are designed to predict $A_{\rm g} \Phi(\alpha)$ (i.e., the phase-dependent brightness), which is subsequently converted to apparent albedo.

The Lambertian surface model adopts Equation~\ref{eqn:lambert} for the Lambert phase function and assumes a gray treatment for the surface flux albedo, $A_{\rm L}$. Ocean glint is simulated using the classical models of \citet{cox&munk1954} and follow a recent revisiting of the expressions by \cite{sayeretal2010}. The glint model is applied over a pixelated, uniform globe observed at a given phase angle. In this case, the observer and solar geometries are known and the real and imaginary indexes of refraction for seawater are adopted. Numerical integration over the illuminated portion of the observable hemisphere then yields a phase-dependent model for the disk-averaged brightness of an ocean world where the free parameter is wind speed over the ocean ($w$; which controls the extent of the wave glitter pattern). Models that include both a Lambertian surface and ocean require an additional parameter that is the ratio of land to ocean ($r_{\rm l/o}$).

Plane-parallel models of the wavelength-dependent, top-of-atmosphere intensity due to Rayleigh, cloud, and optically thin aerosol scattering were generated using a multi-stream, multiple scattering radiative transfer tool\,---\,the Spectral Mapping Atmospheric Radiative Transfer (\texttt{SMART}) model \citep[developed by D.\,Crisp; ][]{meadows&crisp1996}. The \texttt{SMART} model is well-validated \citep[e.g., ][]{crisp1997,schwietermanetal2015,arneyetal2014} and relies on the well-trusted \texttt{DISORT} software \citep{stamnesetal1988} for solutions to the multiple-scattering radiative transfer equation. Mapping the \texttt{SMART}-computed intensities onto a globe and subsequent disk-averaging follows the prescription outlined in \citet{robinson&salvador2022}. The Rayleigh scattering treatment includes surface scattering according to the aforementioned gray surface flux albedo. 

In the cloud model, the medium is taken to be optically thick (so that the solution is insensitive to the underlying surface) and conservatively scattering (as is appropriate for water clouds in visual wavelength range). The aerosol scattering phase function follows a double Henyey-Greenstein phase function \citep{kattawar1975}, 
\begin{equation}
    P \left( \Theta; g_{\rm f},g_{\rm b},f_{\rm f} \right) = f_{\rm f} P_{\rm HG}\left( \Theta; g_{\rm f} \right) + (1-f_{\rm f}) P_{\rm HG}\left( \Theta; g_{\rm b} \right)
\end{equation}
where $\Theta$ is the scattering angle, $g_{\rm f}$ is the asymmetry parameter for the forward-scattered component, $g_{\rm b}$ is the asymmetry parameter for the backward-scattered component (negative, by convention), $f_{\rm f}$ is the fractional weighting of the forward-scattered component, and $P_{\rm HG}$ is the classical Henyey-Greenstein scattering phase function \citep{henyey&greenstein1941},
\begin{equation}
    P_{\rm HG}\left( \Theta; g \right) = \frac{1-g^2}{\left( 1 + g^2 - 2g\cos \Theta \right)^{3/2}} \ .
    \label{eqn:henyey}
\end{equation}
Clouds are assumed to be patchy and cover some fraction of the disk, which introduces a cloud fraction parameter ($f_{\rm c}$).

The optically thin aerosol treatment is intended to distinguish potential false positives for ocean glint through strong forward scattering. The ``thin aerosol'' model component is not intended to represent any single aerosol type but is, instead, a catch-all for scattering from optically thin water clouds (e.g., cirrus), stratospheric sulfuric acid aerosols \citep{jungeetal1961}, and what the Earth sciences call ``clearsky aerosols'' (e.g., dust, soot, smog). Thin aerosol models are assumed to be conservatively scattering, and are executed over a grid of vertical column optical depths spanning optically thin to 10 with a classical Henyey-Greenstein scattering phase function. (Although, at large optical depths the aerosol becomes opaque and behaves more like the cloud model.) Thus, this model has two parameters: the aerosol vertical column optical depth ($\tau_{\rm aer}$) and the aerosol scattering asymmetry parameter ($g_{\rm aer}$).

Regarding gas absorption, as 90\% of the total ozone column mass resides above Earth's troposphere \citep{nicolet1975}, ozone absorption is treated as overlaying the surface, Rayleigh, and cloud models. In this approach, the wavelength-dependent ozone absorption optical depth is used to reduce the direct solar beam incident on, as well as the upwelling intensity from, deeper atmospheric layers.

\added{
The phase dependent models described above predict the disk-averaged intensity given requisite free parameters and certain wavelength-dependent inputs. These spectral quantities are integrated across the visual band (\added{0.4--0.7\,$\upmu$m}; with a solar flux weighting) and converted to apparent albedo using Equation~\ref{eqn:Aapp}. In general, these models have the form,
\begin{equation}
\begin{split}
    \bar{I}\left( \alpha,\lambda \right) = & f_{\rm l} \bar{I}_{\rm l}\left( \alpha, A_{\rm L}, \tau_{\rm O3}, \tau_{\rm R}, \tau_{\rm h} \right) + f_{\rm o} \bar{I}_{\rm o}\left( \alpha, w, \tau_{\rm O3}, \tau_{\rm R}, \tau_{\rm h} \right) \\ & + f_{\rm c} \bar{I}_{\rm c}\left( \alpha, g_{\rm f}, g_{\rm b}, f_{\rm f},\tau_{\rm O3},\tau_{\rm h} \right) + \bar{I}_{\rm h}\left( \alpha, g_{\rm h}, \tau_{\rm h}, ,\tau_{\rm O3} \right) \ ,
\end{split}
\end{equation}
where sub-scripts on the disk-averaged intensity models indicate land, ocean, cloud, and haze contributions, and $\tau_{\rm O3}$ and $\tau_{\rm R}$ are the wavelength-dependent ozone and Rayleigh scattering optical depths, respectively. Models that omit contributions from the land, ocean, or cloud components are achieved by simply setting $f_{\rm l}$, $f_{\rm o}$, and $f_{\rm c}$ equal to zero, respectively. For models with both land and ocean, the land and ocean fractions can be derived given a cloud fraction and ratio of land to ocean via,
\begin{equation}
    f_{\rm o} = \frac{1-f_{\rm c}}{r_{\rm l/o}+1} \ ,
\end{equation}
and
\begin{equation}
    f_{\rm l} = r_{\rm l/o}f_{\rm o} \ .
\end{equation}
}

\subsection{Variability Model}
\begin{figure}
    \centering
    \includegraphics[scale=0.50,trim=0mm 0mm 0mm 0mm]{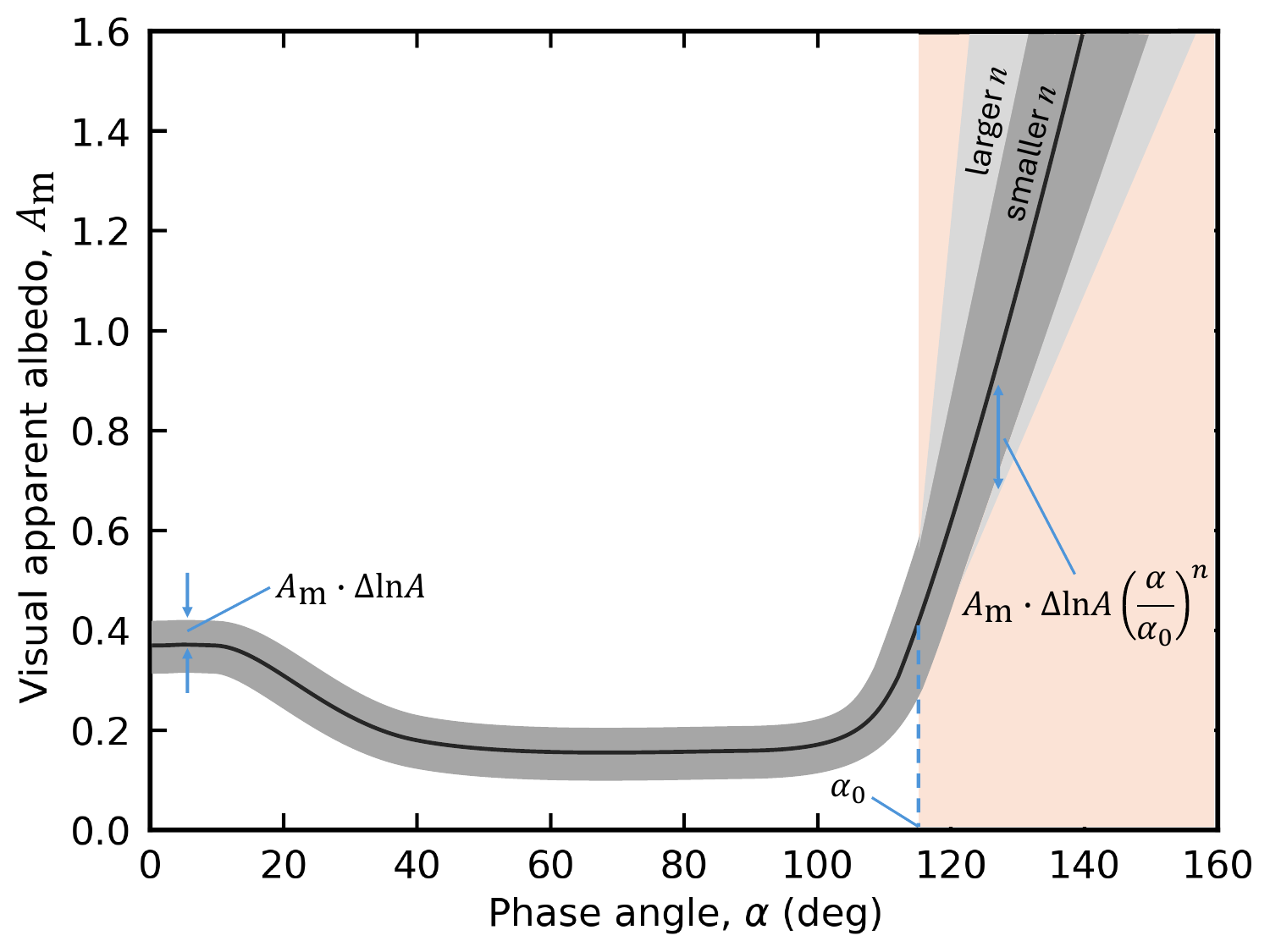}
    \caption{Schematic demonstrating how the adopted variability model parameterizes a variability-driven apparent albedo spread as a function of phase angle. Key parameters are the fractional variability ($\Delta \ln A$), the variability breakpoint phase angle ($\alpha_{0}$), and the high-phase variability power law index ($n$).}
    \label{fig:var_model}
\end{figure}

Figure~\ref{fig:earth_data} demonstrates that Earth's apparent albedo can take on a range of values at a given phase angle. Away from near-full phase, the data in this figure are dominated by Earthshine observations that are taken sporadically and, thus, record \added{variability due to weather, viewing geometry, and seasonal effects.} \added{As an example of these effects, time-varying cloud cover over oceans could introduce strong variability at crescent phases by masking a glint spot.} Initial fits to the phase curve data were performed with a variability standard deviation equal to some fixed fraction of the mean apparent albedo at a given phase angle. However, this approach produced poor fits and is in disagreement with high fidelity Earth models that show increasing fractional variability at crescent phases \citep{robinsonetal2010}. Thus, a three-parameter statistical variability model was adopted that allowed the relative variability to increase beyond some breakpoint phase angle,
\begin{equation}
  \sigma_{\rm m} \left(\alpha; \bm{x} \right) = 
  \begin{cases}
    A_{\rm m}\!\left( \alpha; \bm{x} \right) \Delta \ln A \ , & \alpha \leq \alpha_{0} \\
    A_{\rm m}\!\left( \alpha; \bm{x} \right) \Delta \ln A \cdot \left( \frac{\alpha}{\alpha_{0}} \right)^{\! n} \ , & \alpha > \alpha_{0} \ ,  \\ 
  \end{cases}
  \label{eqn:var_model}
\end{equation}
so that the signal-to-noise ratio ($A_{\rm m}/\sigma_{\rm m}$; \added{recall the definition of the physical model above with an adopted subscript `m'}) is constant at lower phases, decreases as a power law at larger phase angles, and where $\alpha_{0}$ is the breakpoint phase angle, $\Delta \ln A$ is a single value representing the fractional variability in apparent albedo, and $n$ defines the power law. The collection of all fitted physical and variability model parameters is summarized in Table~\ref{tab:params}, and Figure~\ref{fig:var_model} schematically demonstrates the influence of the variability model parameters on the apparent albedo spread as a function of phase angle.

\begin{table}
\centering
\begin{tabular}{lll}
Parameter      & Description & Prior \\ \hline
$A_{\rm L}$    & gray surface flux albedo                       &  $\mathcal{U}\left(0,1\right)$   \\
$w$            & average windspeed over ocean (m/s)             &  $\mathcal{U}\left(0,10\right)$  \\
$g_{\rm f}$    & cloud forward scattering asymmetry parameter   &  $\mathcal{U}\left(0,1\right)$   \\
$g_{\rm b}$    & cloud backward scattering asymmetry parameter  &  $\mathcal{U}\left(-1,0\right)$  \\
$f_{\rm f}$    & cloud forward scattering weight                &  $\mathcal{U}\left(0,1\right)$   \\
$\tau_{\rm h}$ & aerosol vertical column optical depth          &  $\mathcal{U}\left(0,10\right)$  \\
$g_{\rm h}$    & aerosol forward scattering asymmetry parameter &  $\mathcal{U}\left(0,1\right)$   \\
$f_{\rm c}$    & cloud fractional coverage                      &  $\mathcal{U}\left(0,1\right)$   \\
$r_{\rm l/o}$  & ratio of land to ocean coverage                &  $\mathcal{U}\left(0,2\right)$   \\
$\Delta \ln A$ & fractional variability                         &  $\mathcal{U}\left(0,1\right)$   \\
$\alpha_{0}$   & variability breakpoint angle (deg)             &  $\mathcal{U}\left(60,120\right)$\\
$n$            & variability power                              &  $\mathcal{U}\left(0,8\right)$   \\
\end{tabular}
\caption{Physical and variability model parameters as well as adopted priors.}
\label{tab:params}
\end{table}

\subsection{Priors}
Table~\ref{tab:params}, in addition to listing model parameters, gives the adopted prior for each parameter, which, for physical parameters, are generally uninformed and/or physically constrained (e.g., surface flux albedos must be between zero and unity). Exceptions include: (1) limiting the windspeed to less than 15\,m\,s$^{-1}$ to not exceed the range over which the original \citet{cox&munk1954} fits were performed, (2) limiting the thin aerosol optical depth to less than 10 to be within the range of the underlying simulations, and (3) setting an upper-bound to the land-ocean ratio that is safely above realistic values for Earth (which has $r_{\rm l/o}$ of about 0.4). \added{For the three parameters that describe the variability model ($\Delta \ln A$, $\alpha_{0}$, and $n$), uninformed priors were set (via testing) to be sufficiently broad so as to not influence any subsequently inferred parameters.}

Not indicated in Table~\ref{tab:params} are parameters relevant to potential systematic biases in the calibration for the \textit{EPOXI}, \textit{Galileo}, and \textit{LCROSS} observations. These have no impact on the physical model and, as the dataset is dominated by Earthshine and \textit{DSCOVR} data, have limited impact on the analyses that follow. Nevertheless, a parameter describing a systematic offset for the calibrations for each of these missions is included in fits. The prior for this parameter is assumed to be Gaussian in shape with width given by the previously-mentioned systematic calibration uncertainties for these missions (i.e., 0.05, 0.08, and 0.1 for \textit{EPOXI}, \textit{Galileo}, and \textit{LCROSS}, respectively). \added{Fitted values for these calibration offsets are always found to be consistent with their prior, with the \textit{EPOXI} and \textit{Galileo} points found to be biased slightly (i.e., less than 10\%) high when compared to the large number of surrounding Earthshine observations.}

\subsection{Likelihood}
For a given measurement of Earth's visual apparent albedo, $A_{i}$, acquired at a known phase angle, $\alpha_{i}$, and with an associated Gaussian uncertainty, $\sigma_{i}$, the likelihood comes from integrating over the product of a Gaussian distribution for the datapoint and the physical-statistical model in Equation~\ref{eqn:physstatmdl}, yielding,
\begin{equation}
\begin{split}
    \mathcal{L}_{i} & = \int_{0}^{\infty} p_{\rm G}\left(A_{i}|A_{\rm app},\sigma_{i} \right) \cdot p_{\rm G}\left(A_{\rm app} | A_{\rm m}(\alpha_i;\bm{x}), \sigma_{\rm m}(\alpha_i;\bm{x}) \right) dA_{\rm app} \\ & = \int_{0}^{\infty} p_{\rm G}\left(A_{\rm app}|A_{i},\sigma_{i} \right) \cdot p_{\rm G}\left(A_{\rm app} | A_{\rm m}(\alpha_i;\bm{x}), \sigma_{\rm m}(\alpha_i;\bm{x}) \right) dA_{\rm app}\ ,
\end{split}
\label{eqn:likelihood_full}
\end{equation}
\added{where the second equality holds due to the symmetry of the Gaussian distribution}. Via the properties of Gaussian distributions, this expression can be simplified to,
\begin{equation}
    \mathcal{L}_{i} = p_{\rm G}\left(A_{i}|A_{\rm m}(\alpha_i;\bm{x}),\sigma_{i}^{\prime} \right) \ ,
\end{equation}
with,
\begin{equation}
    \sigma_{i}^{\prime} = \sqrt{ \sigma_i^2 + \sigma_{\rm m}^2 } \ .
\end{equation}
\added{Formally this simplified expression only applies if the lower bound in the integral in Equation~\ref{eqn:likelihood_full} is extended to negative infinity. However, this approximation is warranted in what follows as the adopted lower bound of zero is already typically 5--10 standard deviations from the mean, so that effectively zero area exists under the integrand between zero and negative infinity.} The overall likelihood then comes from the product of each $\mathcal{L}_{i}$, with,
\begin{equation}
    \mathcal{L} = \prod_{i} \mathcal{L}_{i} \ .
\end{equation}

\subsection{Example and Validation}
\begin{figure}
    \centering
    \includegraphics[scale=0.45,trim=0mm 0mm 0mm 0mm]{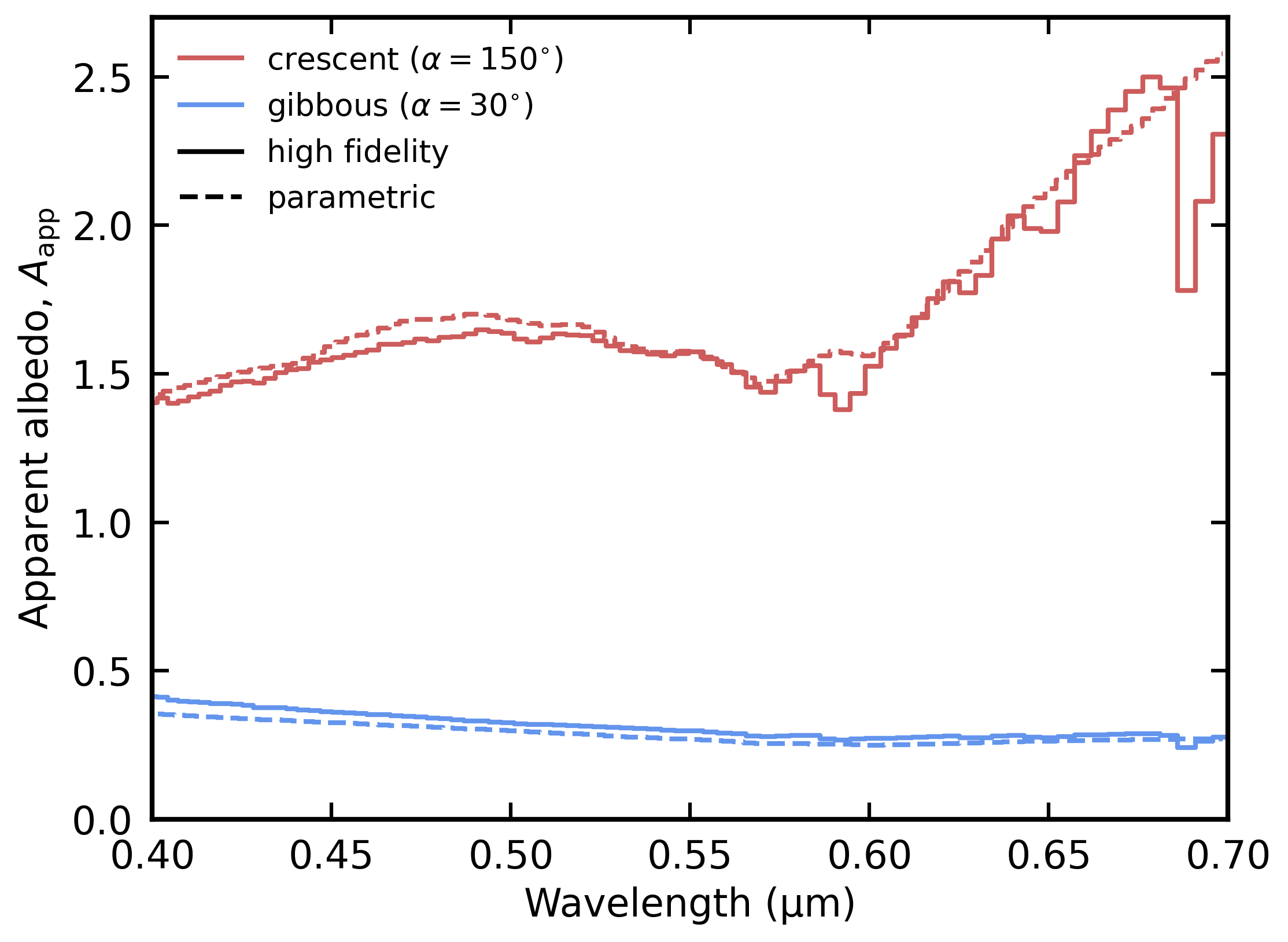}
    \caption{Comparison between the phase- and wavelength-dependent parametric model developed in this work (dashed) to high-fidelity simulated Earth spectra from the Virtual planetary Laboratory Three-Dimensional Spectral Earth Model \citep[solid; data from][]{robinsonetal2010}. Crescent- (red) and gibbous-phase (blue) spectra are shown, and the spectral range highlights the visual band. The simple model sufficiently reproduces results from the high-fidelity tool.}
    \label{fig:valid}
\end{figure}

\begin{figure}
    \centering
    \includegraphics[scale=0.75,trim=0mm 0mm 0mm 0mm]{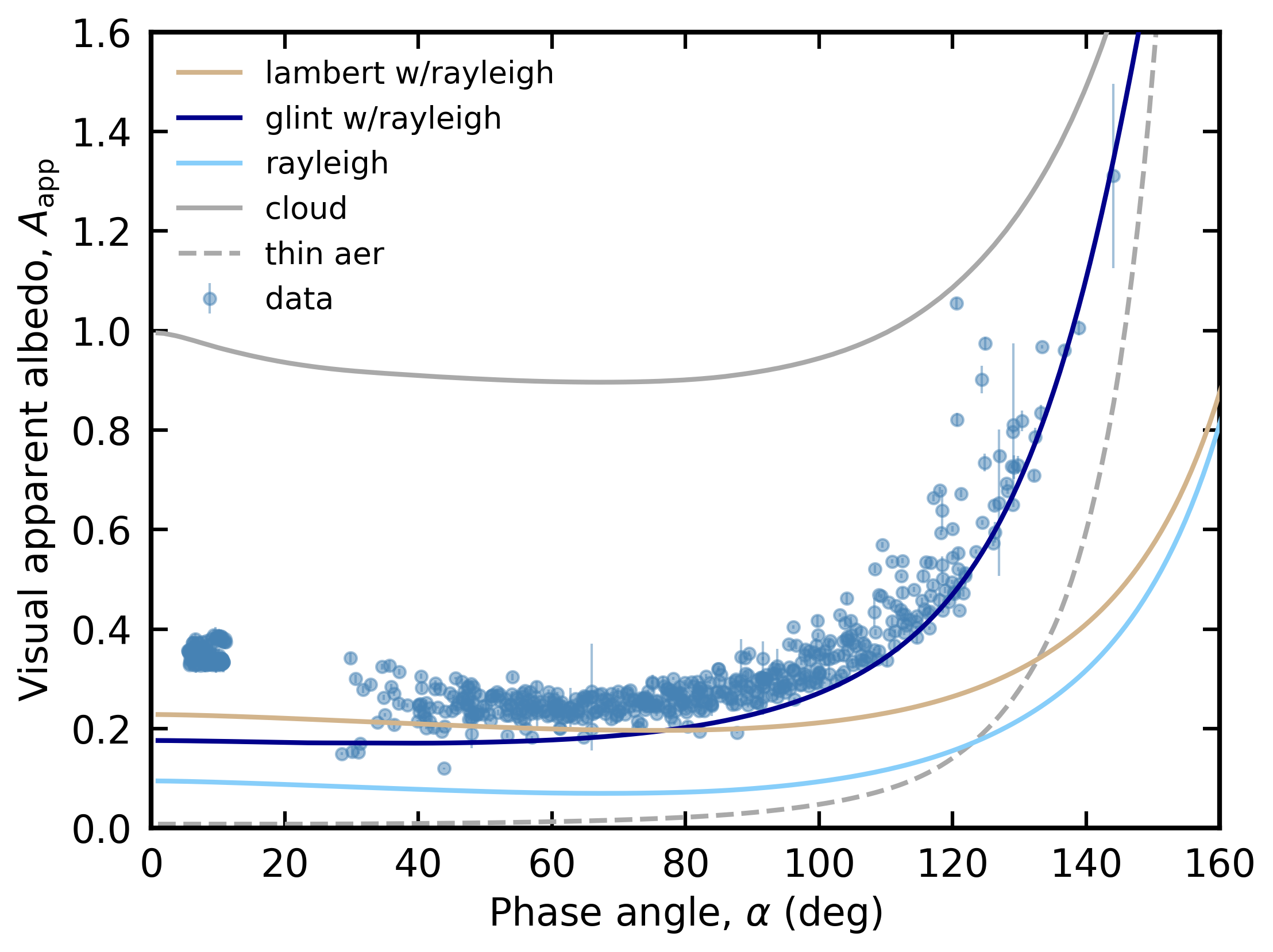}
    \caption{Visual band phase curves, shown as apparent albedo, demonstrating the phase-dependence of the various components of the model developed in this work. Earth phase curve data from Figure~\ref{fig:earth_data} are shown as points for comparison. Component phase curves do not include a spatial weighting for coverage on the disk, and the Rayleigh curve is for Rayleigh scattered light that has not been also reflected by the surface.}
    \label{fig:demo}
\end{figure}

The model described above is minimally parametric and produces phase- and wavelength-dependent disk-averaged brightness that, when integrated over the visual band \added{(0.4--0.7\,$\upmu$m)}, can be fit to the visual phase curve data described in Section~\ref{sec:data}. Figure~\ref{fig:valid} compares apparent albedo spectra from the new tool to equivalent outputs from the high-fidelity (and computationally complex) Virtual Planetary Laboratory Three-Dimensional Spectral Earth Model \citep{tinettietal2006a,tinettietal2006b,robinsonetal2010,robinsonetal2011,schwietermanetal2015}. Parameters for the simpler model are set to known values for Earth, with 20\% fractional coverage of optically thick water clouds and a weakly forward scattering thin aerosol component ($\tau_{\rm aer}$ and $g_{\rm aer}$ of 0.1 and 0.2, respectively). The simple model well reproduces spectra from the high-fidelity tool, excepting the intentional omission of lower-importance molecular oxygen absorption at 0.63\,$\upmu$m and 0.69\,$\upmu$m and water vapor opacity at 0.59\,$\upmu$m and 0.65\,$\upmu$m. \added{Additionally, the simple model does not perfectly reproduce Earth's color owing to its gray surface flux albedo treatment, which cannot capture wavelength-dependent contributions due to, for example, vegetation and land (that cover a minority of Earth's surface).} Finally, as an example, Figure~\ref{fig:demo} demonstrates the shapes of the visual band phase curves generated by the different components used in the validation simulation. Contributions are unweighted (e.g., the ``cloud'' contribution in the demonstration figure would be weighted by a fractional cloudiness in a complete model) and the Rayleigh curve shows contributions from Rayleigh scattering when surface reflectance is removed.

%%%
%
\section{Results} \label{sec:results}
%
%%%

A suite of models were fit to the visual phase curve data shown in Figure~\ref{fig:earth_data} with the goal of identifying which model components are required for best-explaining Earth's observed phase curve. From the best-explaining models, a most-physical model is then used to derive key planetary properties (with uncertainties) for Earth, including its visual geometric albedo and its visual spherical albedo. Finally, a simple analytic expression is fit to the phase curve data to produce a model that others can use to easily reproduce Earth's phase curve to a sufficient degree of accuracy.

Fits were performed using the \texttt{dynesty} dynamic nested sampling package \citep{speagle2020,koposovetal2024}, \added{where nested sampling is an approach to Markov chain Monte Carlo that both draws samples from the prior distribution (to improve efficiency) and groups samples in such a way to more accurately evaluate integrals over parameter space}. Critically, \texttt{dynesty} adopts published approaches for sampling \citep{skilling2006}, nested sampling \citep{skilling2004}, dynamic nested sampling \citep{higsonetal2019}, and bounding \citep{buchner2016,buchner2019}. All fits used 1,000 live points to thoroughly explore the posterior distribution. Posteriors were visualized using the \texttt{corner} package \citep{foremanmackey2016}.

\subsection{Grid of Model Fits}

Table~\ref{tab:models} details a grid of models that were fit to the Earth visual phase curve data. Excepting one cloud-free model, the grid was assembled by requiring at least one surface component (either Lambertian and/or ocean glint) and the optically thick cloud component. From the perspective of ``Earth as an exoplanet,'' retrievals on optical spectra of Earth generally detect both surface and thick cloud contributions \citep{fengetal2018}, so these components are reasonable to include when performing general fits to Earth's phase curve. \added{Table~\ref{tab:modelpars} contains a more detailed listing of the physical parameters adopted in the models identified in Table~\ref{tab:models}. Not listed in Table~\ref{tab:modelpars}, due to their commonality across all models, are the three parameters in the variability model ($\Delta \ln A$, $\alpha_{0}$, and $n$) and the three parameters used to treat systematic calibration biases for the \textit{EPOXI}, \textit{Galileo}, and \textit{LCROSS} observations.}

\begin{figure}
    \centering
    \includegraphics[scale=0.3,trim=0mm 0mm 0mm 0mm]{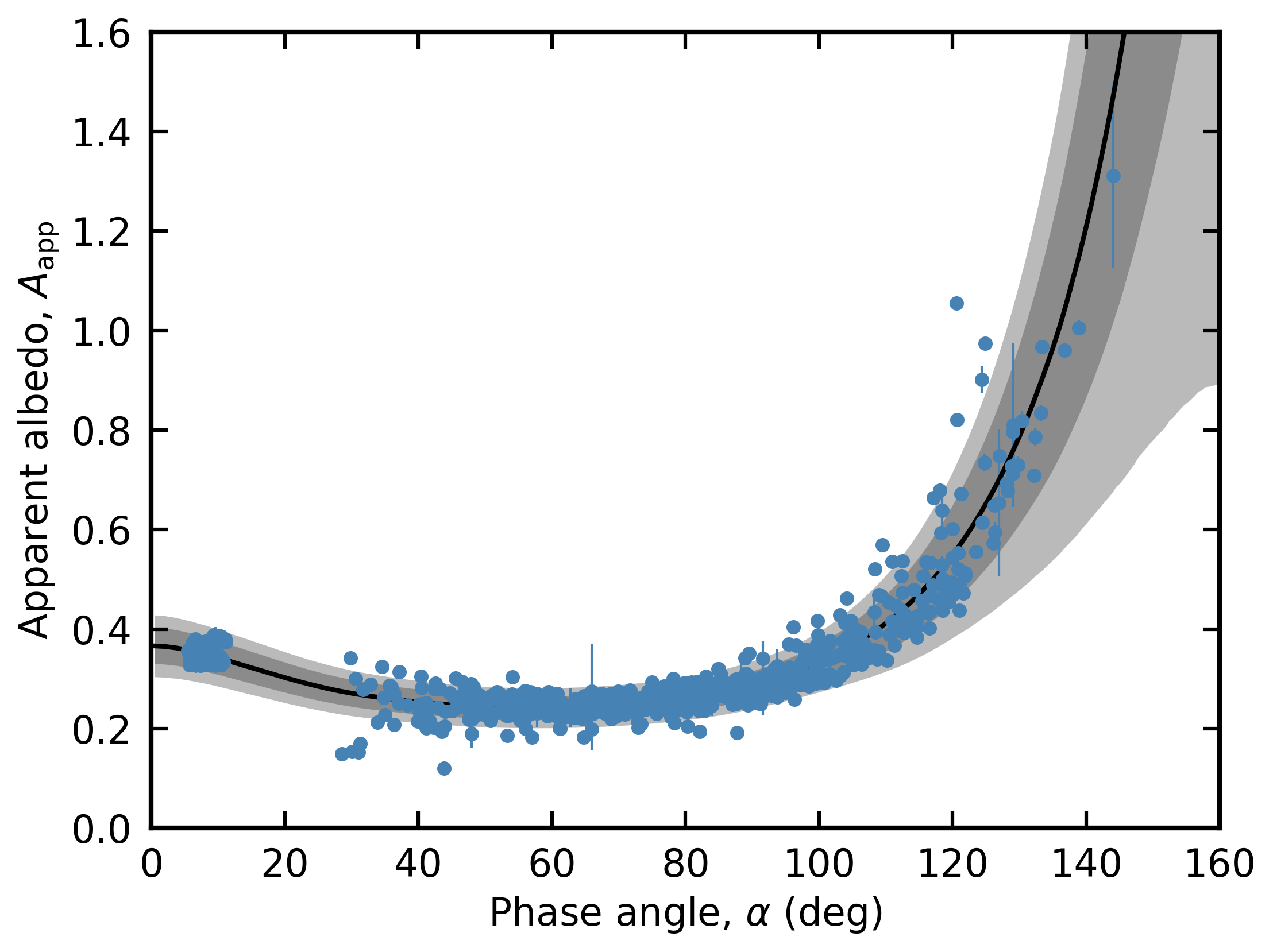}
    \includegraphics[scale=0.3,trim=0mm 0mm 0mm 0mm]{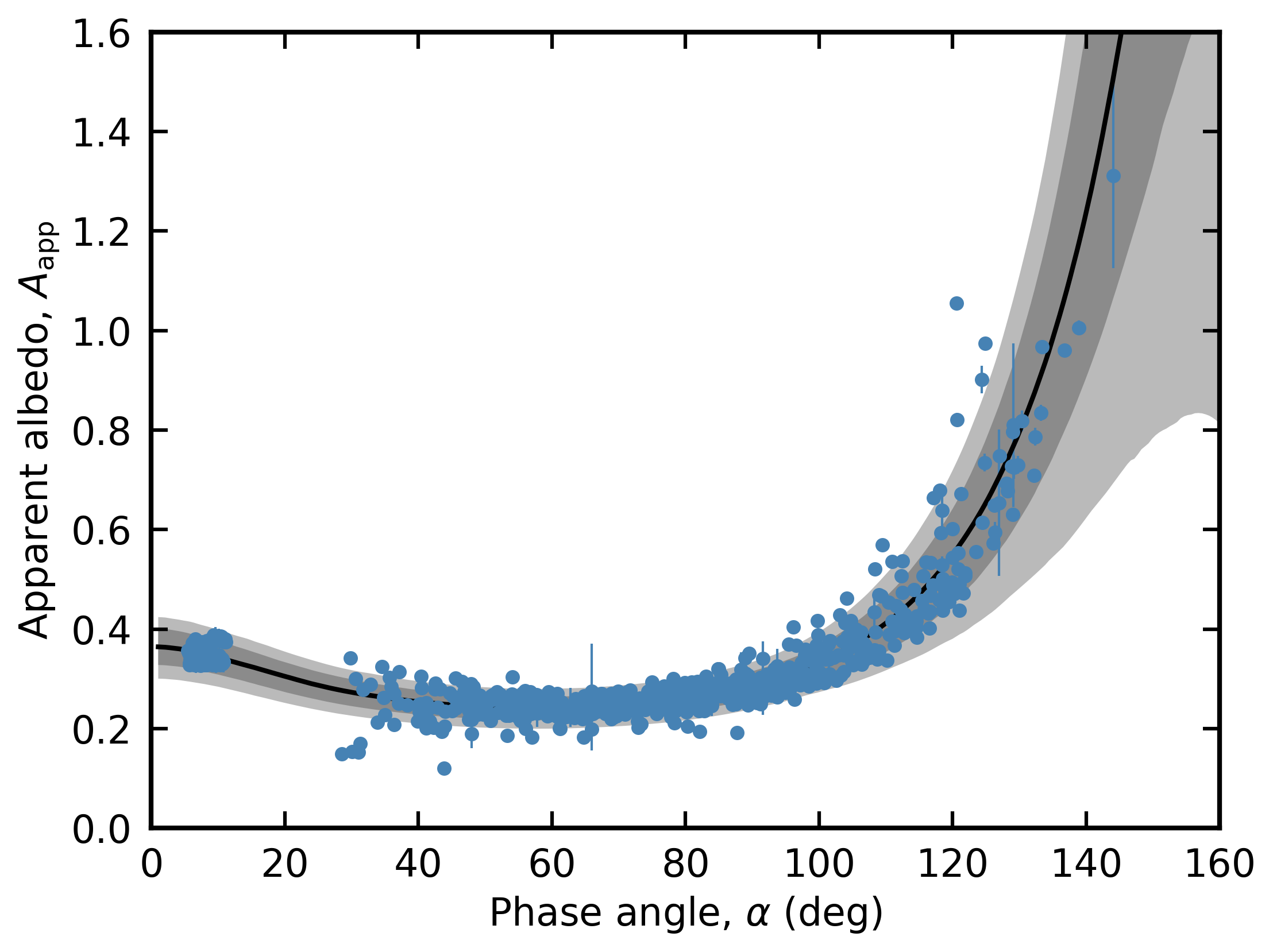}
    \includegraphics[scale=0.3,trim=0mm 0mm 0mm 0mm]{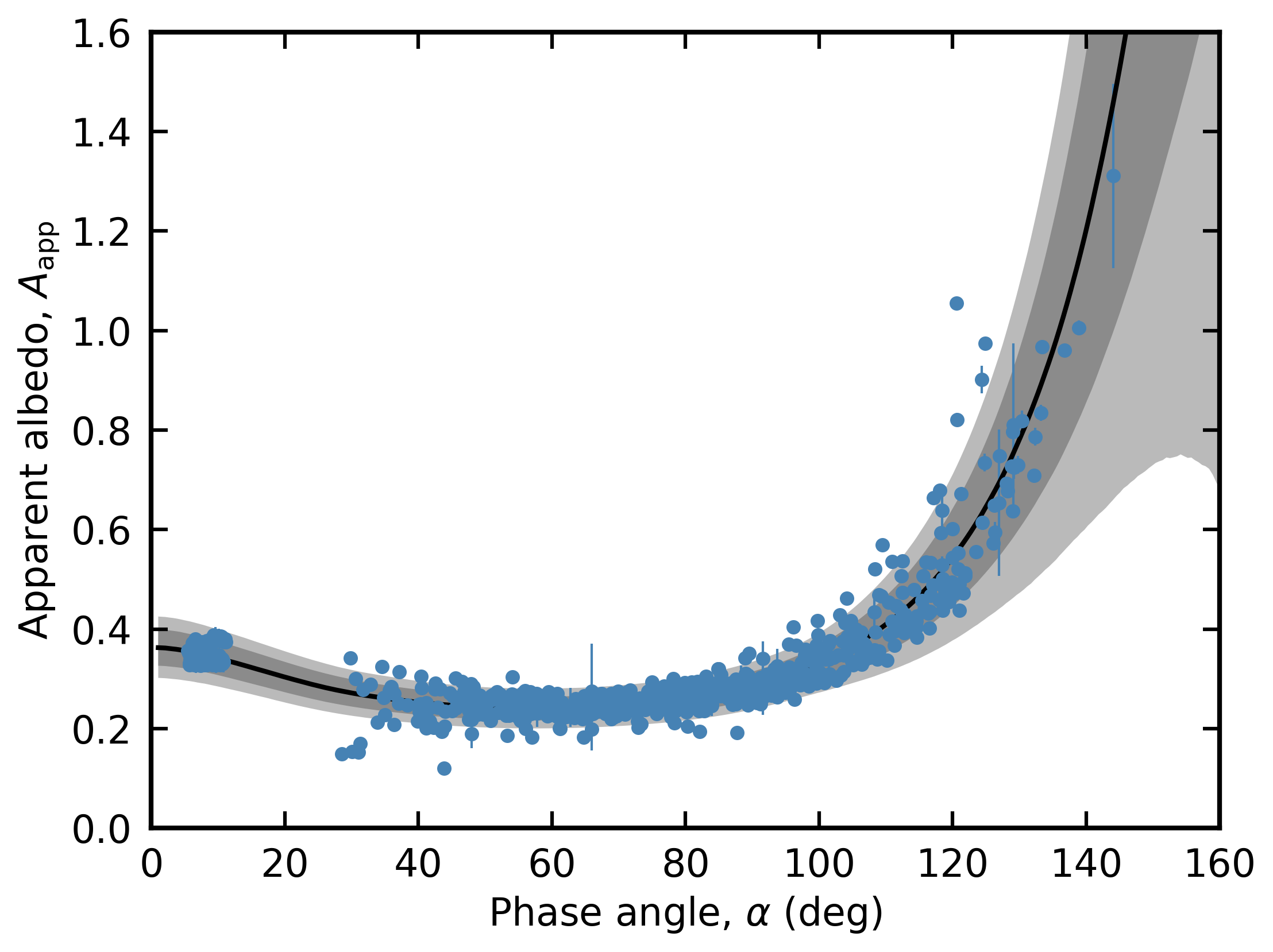}
    \caption{Data-model comparisons for best-fitting models from case numbers 02 (left), 05 (middle), and 07 (right), with components as given in Table~\ref{tab:models}. All models reproduce the phase curve observations well with reduced chi-squared values very near to unity. \added{As a reminder, the geometric albedo is equal to 2/3 of the apparent albedo at full phase.}}
    \label{fig:bestfits}
\end{figure}

The difference in the log-Bayesian evidence ($\ln \mathcal{Z}$) for any pair of models in Table~\ref{tab:models} yields the log-Bayes factor for model selection purposes. Log-Bayes factors greater than roughly 5 are generally interpreted as ``strong evidence'' for the model with the larger evidence whereas log-Bayes factors of one or less are inconclusive \citep{trotta2008,thorngrenetal2025}. Thus, Models 02, 05, and 07 best-explain the Earth visual phase curve observations. Best-fitting models from each of these cases are shown in Figure~\ref{fig:bestfits}, which all have reduced chi-squared values very near to unity. For completeness, Appendix~\ref{sec:corners} provides the corner plots for Models 02, 05, and 07. \added{As a check, a fit was performed that omitted any surface contribution and only used planet-wide thick cloud and haze components; this model struggled to match the brightness scale and shape of the phase-dependent Earth observations and obtained a log-Bayesian evidence of $-96$.}

\begin{table}
\centering
\begin{tabular}{cccccc}
               & \multicolumn{4}{c}{Component Included} &                   \\
Model No.      & Lambert & Glint & Cloud & Thin Aer.    & $\ln \mathcal{Z}$ \\ \hline
 01            & Y       & N     & Y     & N            &  1041             \\
 02            & Y       & N     & Y     & Y            &  1188             \\
 03            & Y       & N     & N     & Y            &  994              \\
 04            & N       & Y     & Y     & N            &  1174             \\
 05            & N       & Y     & Y     & Y            &  1188             \\
 06            & Y       & Y     & Y     & N            &  1172             \\
 07            & Y       & Y     & Y     & Y            &  1187             \\
\end{tabular}
\caption{Modeling grid setup and resulting log-Bayesian evidence.}
\label{tab:models}
\end{table}

\begin{table}
\centering
\begin{tabular}{cccccccccc}
          & \multicolumn{9}{c}{Parameter Included} \\
Model No. & $A_{\rm L}$ & $w$ & $g_{\rm f}$ & $g_{\rm b}$ & $f_{\rm f}$ & $\tau_{\rm h}$ & $g_{\rm h}$ & $f_{\rm c}$ & $r_{\rm l/o}$ \\ \hline
 01       & Y           & N   & Y           & Y           & Y           & N              & N           & Y           & N             \\
 02       & Y           & N   & Y           & Y           & Y           & Y              & Y           & Y           & N             \\
 03       & Y           & N   & N           & N           & N           & Y              & Y           & N           & N             \\
 04       & N           & Y   & Y           & Y           & Y           & N              & N           & Y           & N             \\
 05       & N           & Y   & Y           & Y           & Y           & Y              & Y           & Y           & N             \\
 06       & Y           & Y   & Y           & Y           & Y           & N              & N           & Y           & Y             \\
 07       & Y           & Y   & Y           & Y           & Y           & Y              & Y           & Y           & Y             \\
\end{tabular}
\caption{Physical parameters adopted in models.}
\label{tab:modelpars}
\end{table}

\subsection{Earth Planetary Properties}

The model that includes a Lambertian surface contribution, oceans, optically thick clouds, and optically thin aerosols (i.e., Model 07) is one of the three best-fitting models and, from \textit{a priori} experience, the most physically appropriate model of these for Earth. Further analysis of this model can yield key reflectance-related quantities for Earth with associated uncertainties. However, the reported reflectance-related quantities for Earth are generally insensitive to whether Model 02, 05, or 07 are adopted, as these models all fit the observations (and perform) quite similarly.

\begin{figure}
    \centering
    \includegraphics[scale=0.45,trim=0mm 0mm 0mm 0mm]{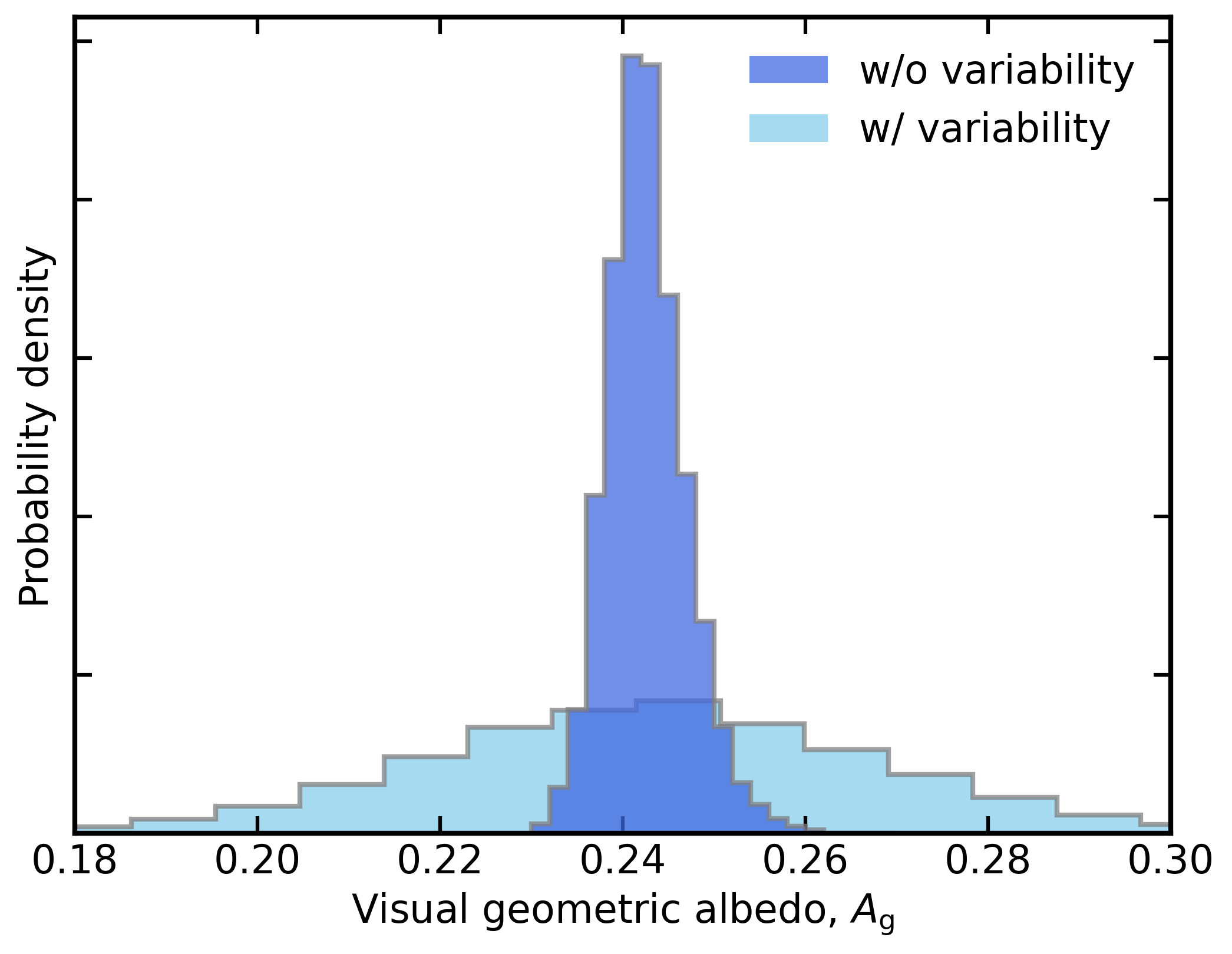}
    \caption{Constraints on Earth's visual geometric albedo. The distribution with variability represents the spread in full-phase brightness that would obtained from repeat measurements. The distribution without variability indicates the true constraint on the geometric albedo and represents the confidence with which the center (median) of the full-phase brightness distribution is known.}
    \label{fig:geom_alb}
\end{figure}

\begin{figure}
    \centering
    \includegraphics[scale=0.45,trim=0mm 0mm 0mm 0mm]{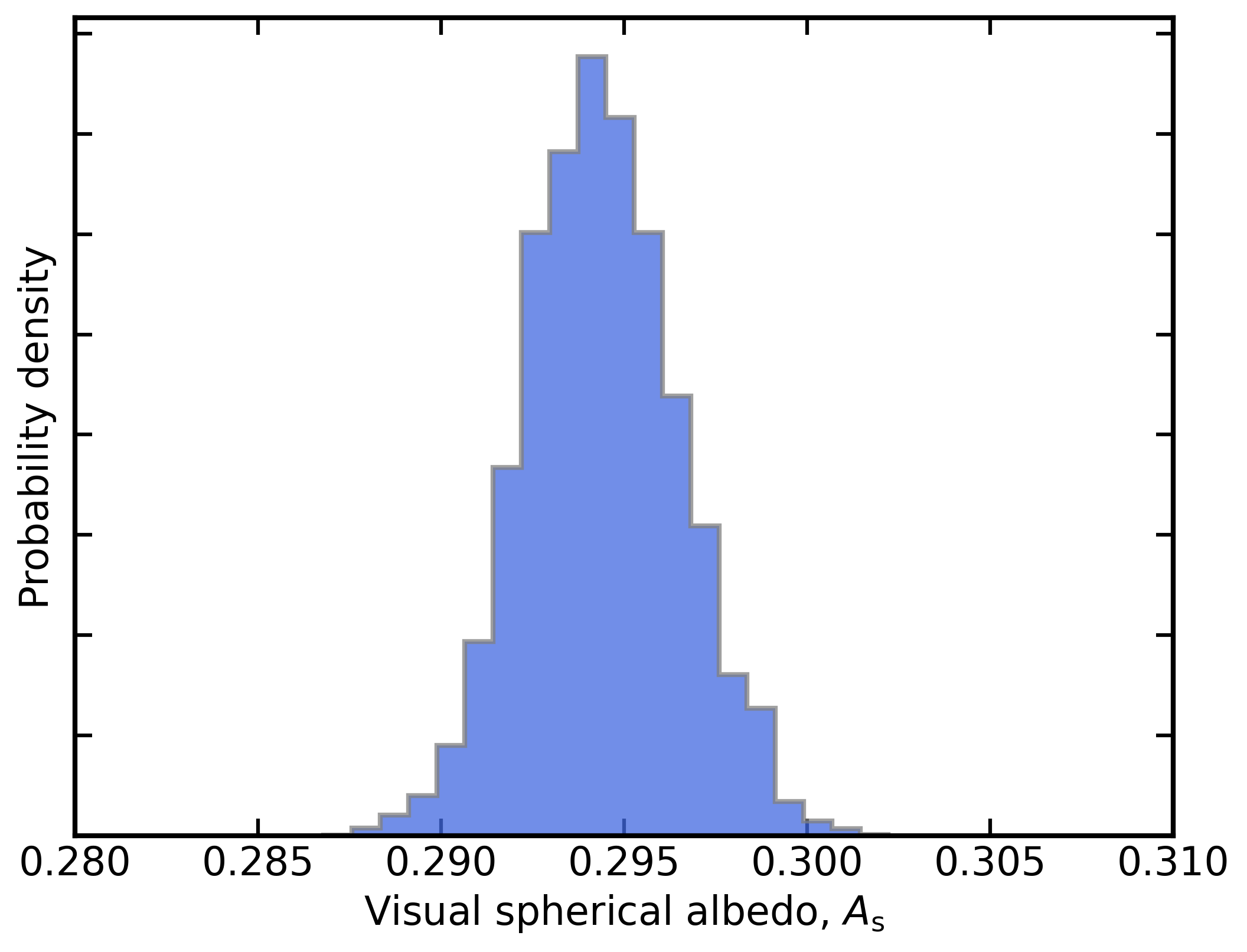}
    \includegraphics[scale=0.45,trim=0mm 0mm 0mm 0mm]{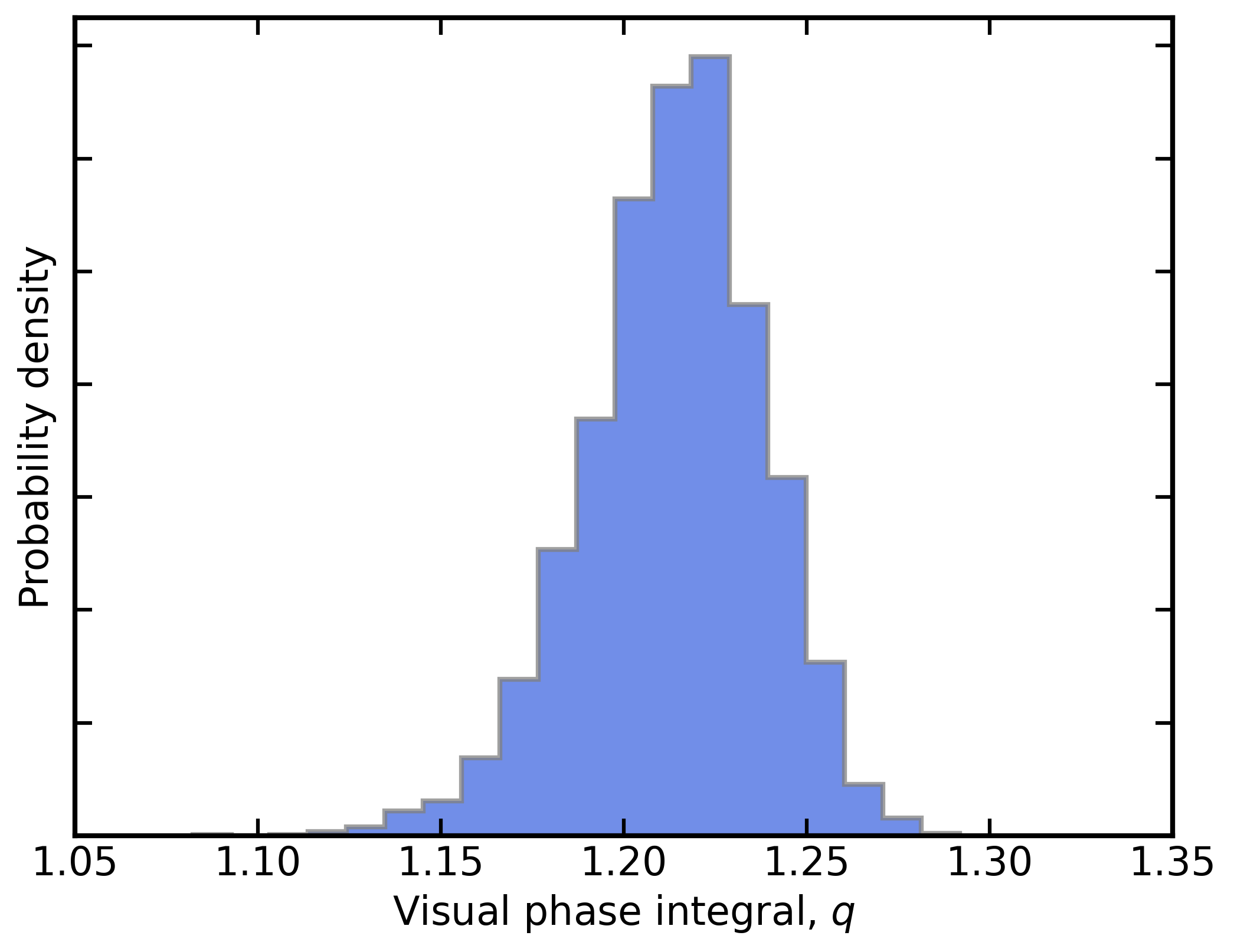}
    \caption{Constraints on Earth's visual spherical albedo (left) and phase integral (right).}
    \label{fig:earth_props}
\end{figure}

Figure~\ref{fig:geom_alb} shows derived constraints on Earth's visual geometric albedo. Two distributions are shown, one that includes the variability effects from the physical-statistical model and one that excludes the variability effects (i.e., just the physical model). The distribution with variability indicates the spread in geometric albedos that would be expected from an experiment that repeatedly measures Earth's rotationally-averaged full-phase visual brightness. The distribution without variability indicates the ``true'' visual geometric albedo on top of which weather introduces variability. In effect, the probability distribution for this true visual geometric albedo indicates the confidence in knowing the center of the variability-affected brightness distribution at full phase. Earth's visual geometric albedo is inferred to be $0.242^{+0.005}_{-0.004}$ for the 16/50/84${\rm th}$ percentiles (adopted hereafter; \added{analogous to $+/-1\sigma$ for a Gaussian distribution}). Integrating a large statistical sample of individual phase curve models over phase angle yields a visual spherical albedo (Equation~\ref{eqn:sphere_alb}) of  $0.294^{+0.002}_{-0.002}$ and a phase integral (Equation~\ref{eqn:phaseint}) of $1.22^{+0.02}_{-0.03}$. Figure~\ref{fig:earth_props} shows the posterior distributions that are the constraints on Earth's spherical albedo and phase integral. Finally, as the underlying physical-statistical model is spectrally dependent, constraints derived on Earth's geometric albedo in B (0.4--0.5\,$\upmu$m), V (0.5--0.6\,$\upmu$m), and R (0.6--0.7\,$\upmu$m) bands are, respectively, $0.277^{+0.005}_{-0.004}$, $0.226^{+0.004}_{-0.004}$, and $0.221^{+0.004}_{-0.004}$.

\begin{figure}
    \centering
    \includegraphics[scale=0.75,trim=0mm 0mm 0mm 0mm]{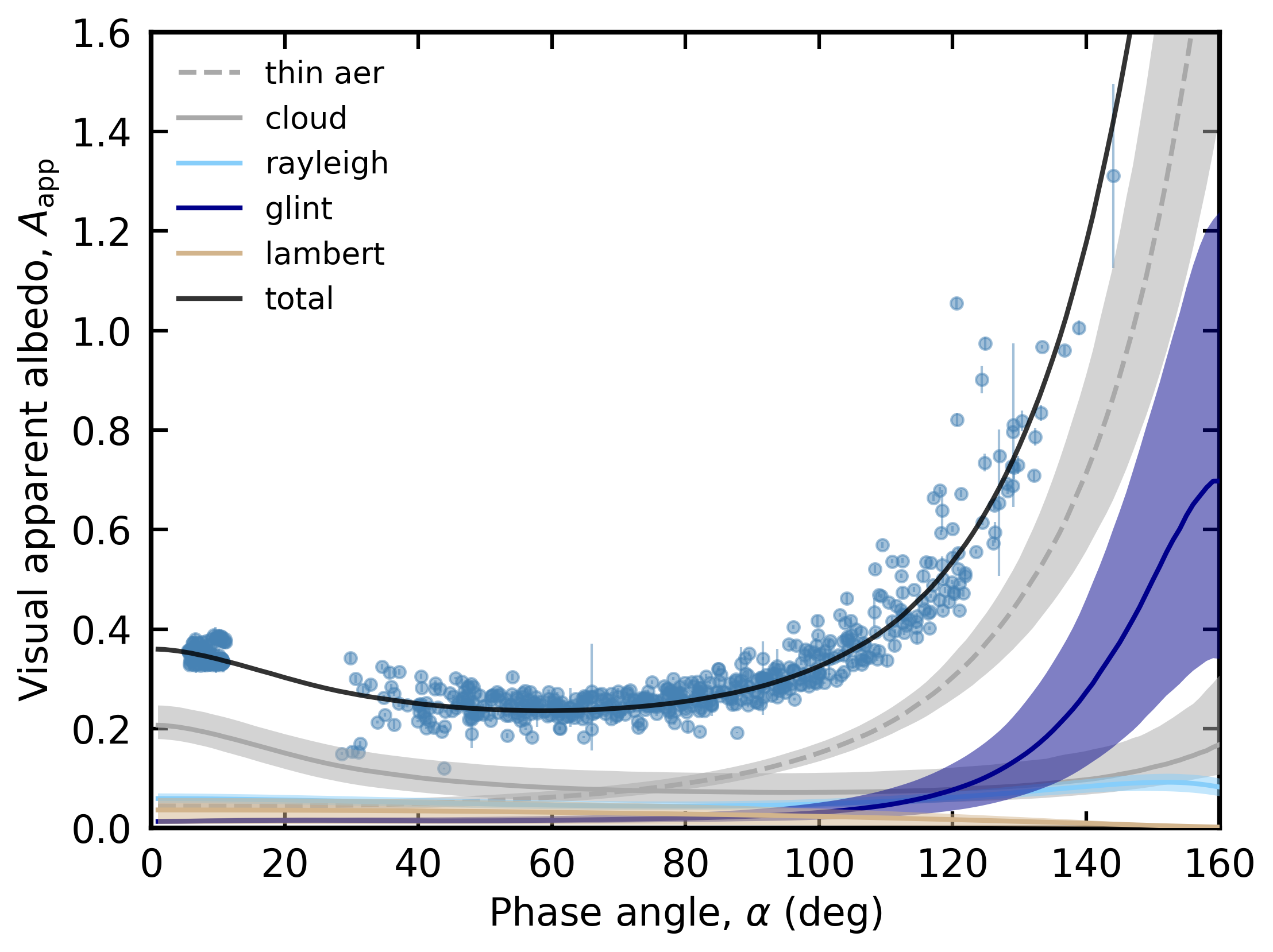}
    \caption{Contributions from different model components to Earth's visual phase-dependent brightness. Median contributions are shown with a line (where color or style indicates the model component) and 16/84\% confidence intervals (i.e., $1\sigma$ for a Gaussian distribution) are indicated by associated swaths. Observational data are also shown. \added{As a reminder, the geometric albedo is equal to 2/3 of the apparent albedo at full phase.}}
    \label{fig:process_swaths}
\end{figure}

\begin{figure}
    \centering
    \includegraphics[scale=0.75,trim=0mm 0mm 0mm 0mm]{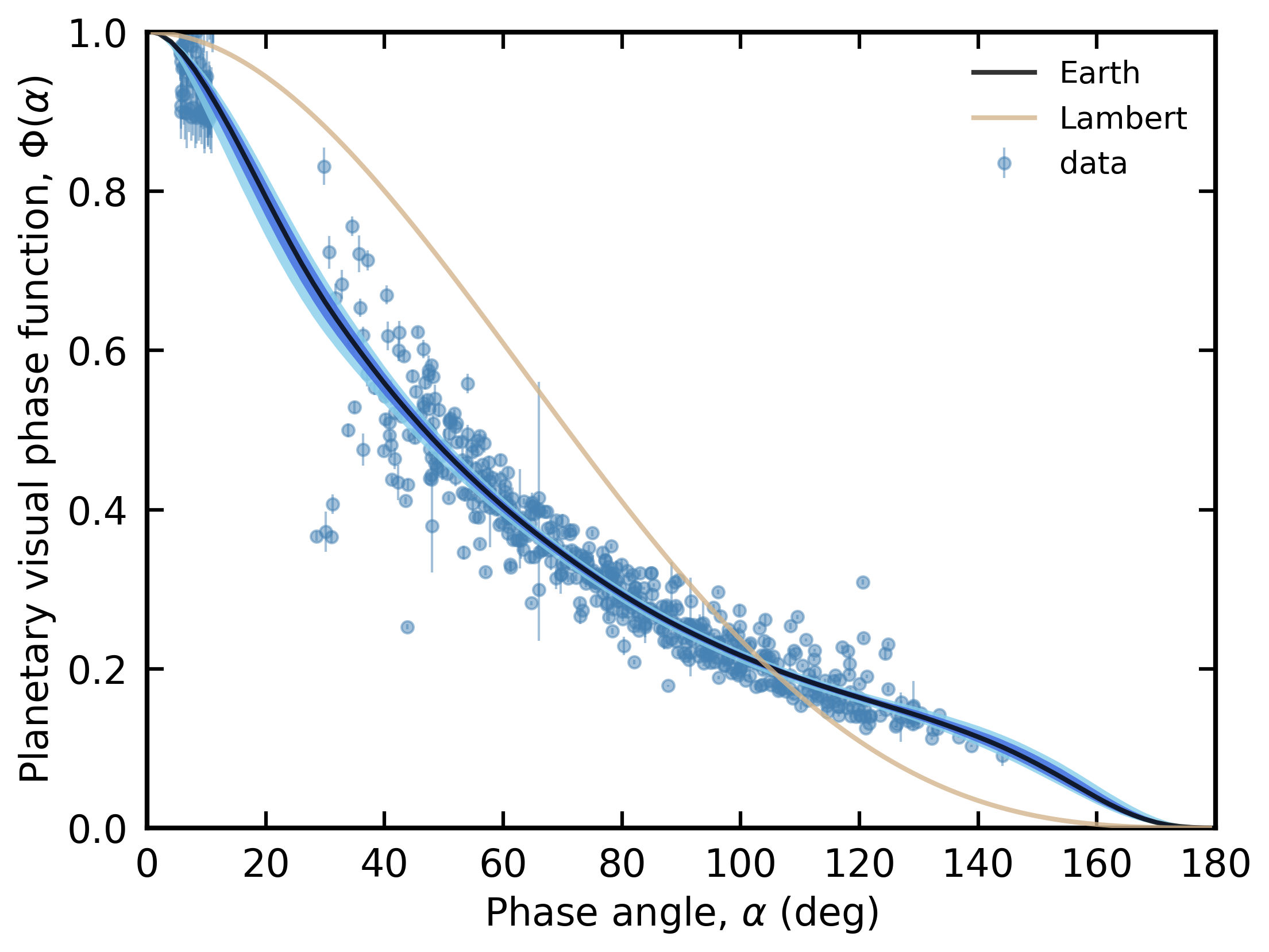}
    \caption{Constraints on Earth's visual phase function. Observational data and the Lambert phase function are shown. The 2.3/16/50/84/97.7\% confidence intervals (i.e., $1\sigma$ and $2\sigma$ for a Gaussian distribution) are indicated in dark and light blue swaths.}
    \label{fig:earth_phase}
\end{figure}

Further applications of Model 07, constrained by the observations, yields constraints on other fundamental planetary properties for Earth. Figure~\ref{fig:earth_phase} shows constraints on Earth's phase function alongside a Lambertian phase function. The former deviates substantially from the latter. Figure~\ref{fig:process_swaths} separates out the contributions to Earth's phase-dependent visual brightness due to the processes included in the physical model. Thick clouds dominate the brightness near full phase while thin aerosol forward scattering and glint dominate at extreme crescent phases.

\begin{figure}
    \centering
    \includegraphics[scale=0.45,trim=0mm 0mm 0mm 0mm]{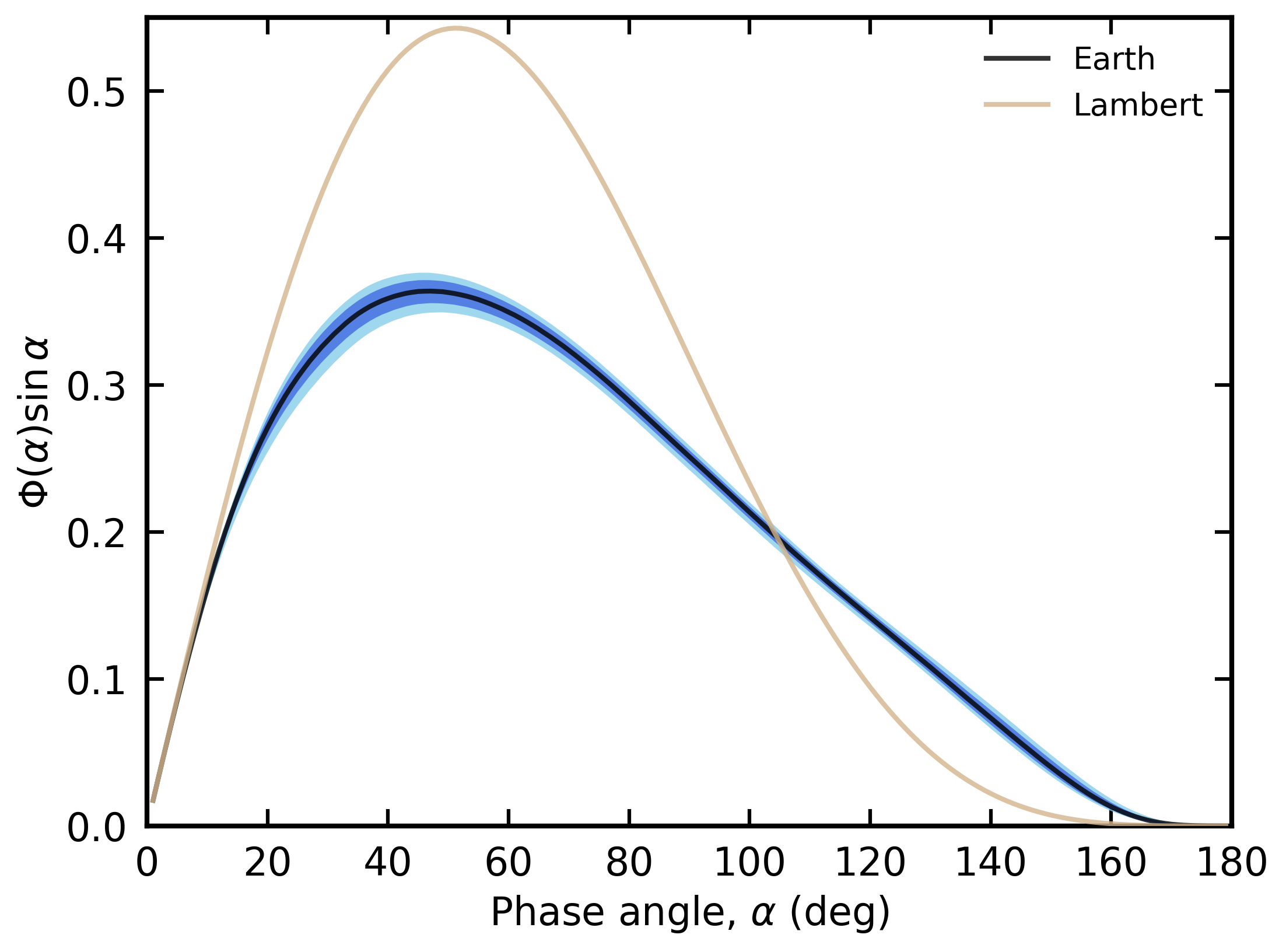}
    \caption{Constraints on the product $\Phi (\alpha) \sin \alpha$, which indicates phases that more-strongly contribute to Earth's visual spherical albedo. The 2.3/16/50/84/97.7\% confidence intervals (i.e., $1\sigma$ and $2\sigma$ for a Gaussian distribution) are indicated in dark and light blue swaths.}
    \label{fig:earth_phaseint}
\end{figure}

\begin{figure}
    \centering
    \includegraphics[scale=0.45,trim=0mm 0mm 0mm 0mm]{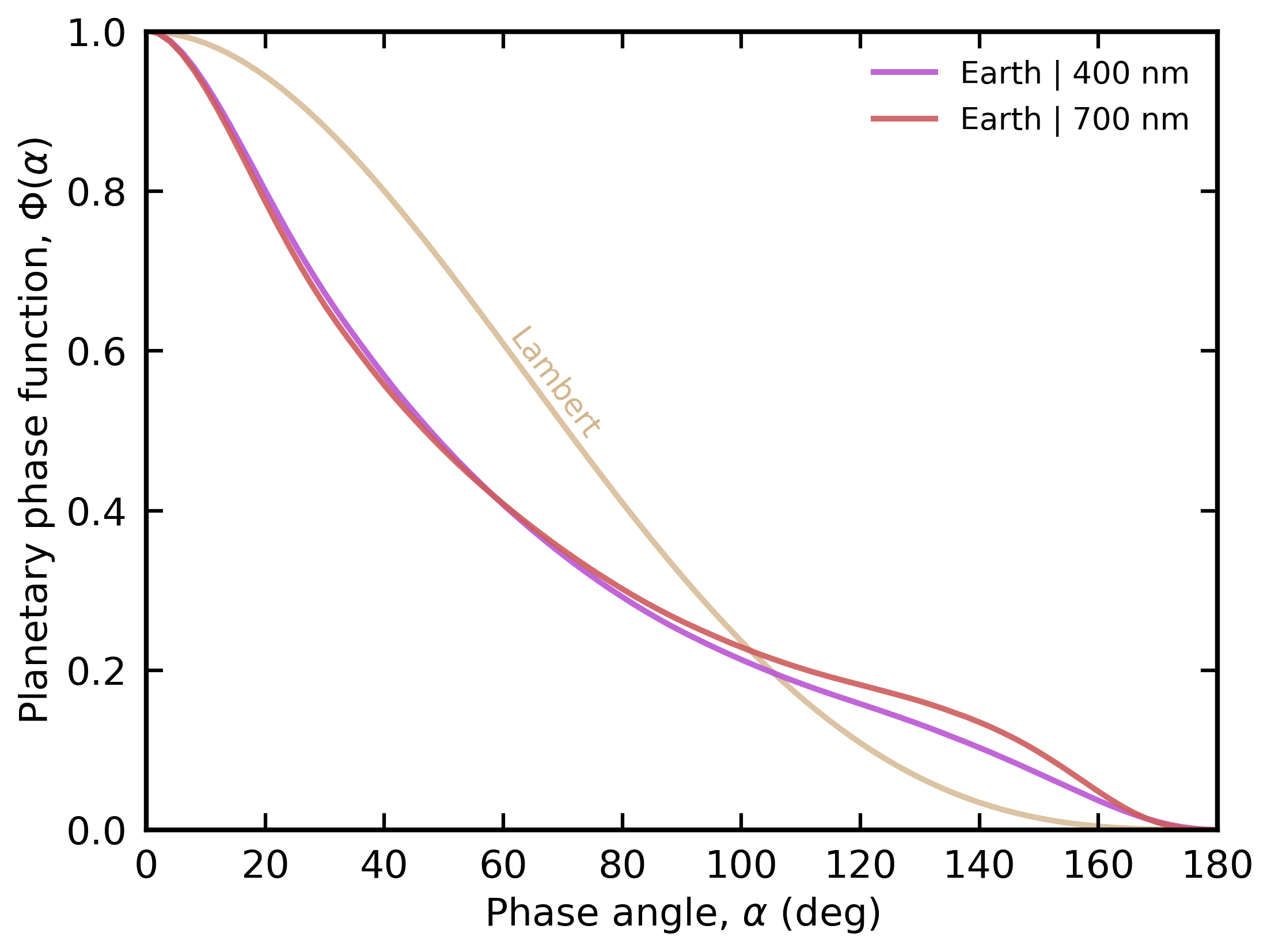}
    \caption{Median constraints on Earth's phase function at different wavelengths. The Lambert phase function is also shown for comparison.}
    \label{fig:earth_phase_color}
\end{figure}

Finally, Figure~\ref{fig:earth_phaseint} shows constraints on the product of the phase function with $\sin \alpha$. As discussed by \citet{garciamunozetal2017}, and as is apparent from the integrand in Equation~\ref{eqn:sphere_alb_simp}, this product highlights the phases that contribute most strongly to the spherical albedo and, thus, play a more important role in planetary energy balance. Finally, as the underlying and constrained model contains a spectral treatment, Figure~\ref{fig:earth_phase_color} shows the median inferred phase function at violet and red wavelengths.

\subsection{A Happy Accident: An Analytic Expression of Earth's Visual Phase Function}

\begin{figure}
    \centering
    \includegraphics[scale=0.45,trim=0mm 0mm 0mm 0mm]{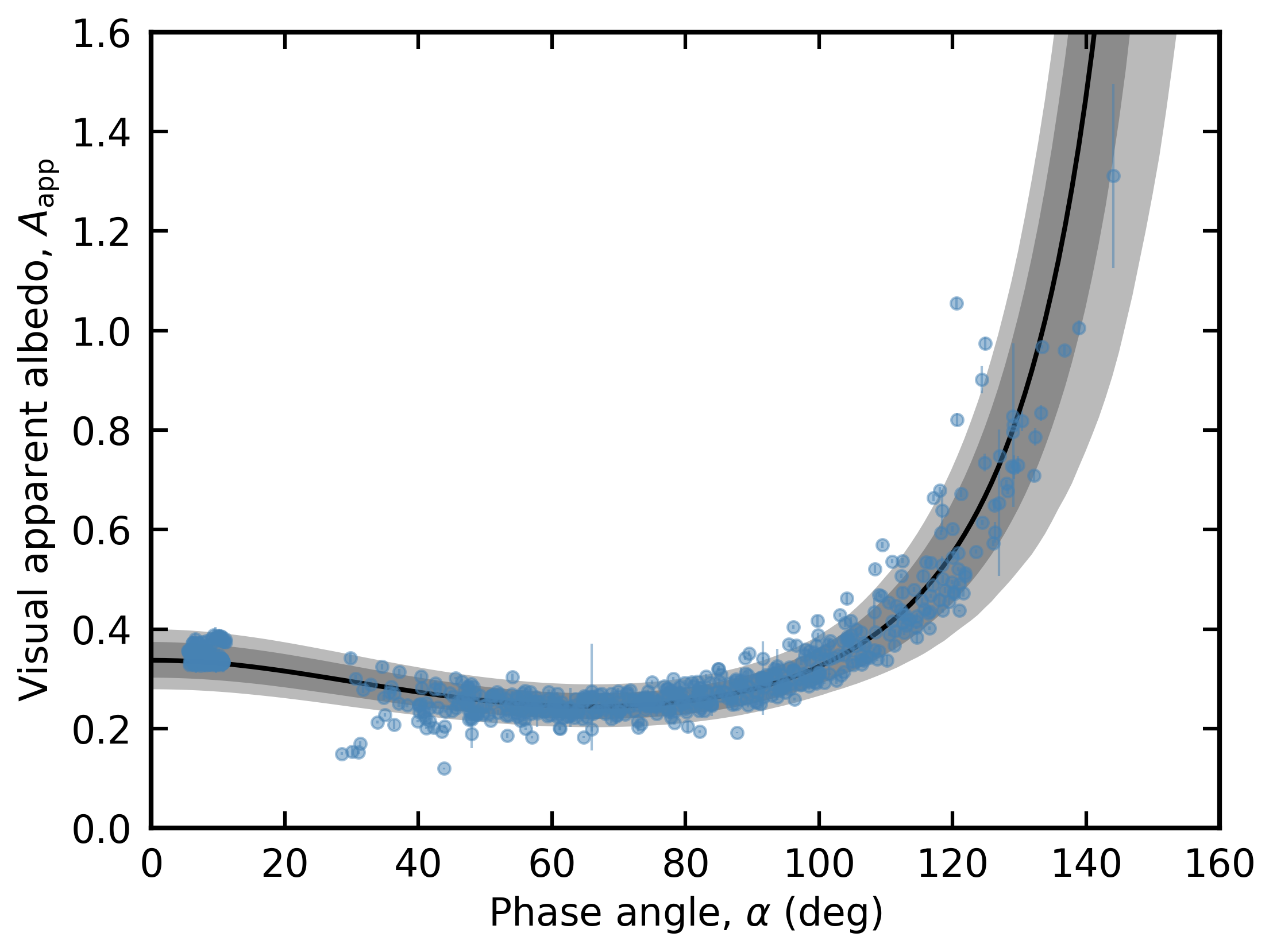}
    \caption{Data-model comparison for the simple analytic (``happy accident'') model, where a simple analytic expression provides a strong reproduction of Earth's visual phase curve. \added{As a reminder, the geometric albedo is equal to 2/3 of the apparent albedo at full phase.}}
    \label{fig:happy_acc}
\end{figure}

An error early in the research process inadvertently led to directly fitting a Henyey-Greenstein phase function normalized at zero degrees to Earth's visual phase curve data, rather than fitting with the physical-statistical models described above. This analytic model was of the form,
\begin{equation}
    \frac{ \bar{I}\left (\alpha \right) }{F_{\rm s}} \approx \frac{f}{\pi} \frac{ P_{\rm HG} \left( \Theta(\alpha); g \right ) }{ P_{\rm HG} \left( \Theta(0^{\circ}); g \right ) } \ ,
    \label{eqn:happyacc}
\end{equation}
where a normalization factor, $f$, and asymmetry parameter, $g$, were fitted parameters, \added{and different scattering direction conventions lead to an angle conversion of $\Theta=180^{\circ}-\alpha$}. \added{Earlier works \citep{brunaetal2023} have represented a planetary phase function using the Henyey-Greenstein scattering phase function, but the novel normalization adopted here esnures that the phase function component in Equation~\ref{eqn:happyacc} is equal to unity at full phase.} \added{The fit also included the previously-described (Equation~\ref{eqn:var_model}) variability model parameters, which are the fractional variability ($\Delta \ln A$), the variability breakpoint phase angle ($\alpha_{0}$), and the high-phase variability power law index ($n$).} Figure~\ref{fig:happy_acc} compares the results from this simple-model fit to the phase-dependent visual apparent albedo data. The best-fit model from this exercise has $f=0.23$, $g=-0.33$, $\Delta \ln A = 0.11$, $\alpha_0=110^{\circ}$, and $n=4.6$. The reduced chi-squared for this best-fit model is 0.96, indicating that this simple approach produces a reliable analytic stand-in for Earth's true visual phase curve.

In this ``happy accident'' model, the parameter $f$ is also equal to the geometric albedo. However, the best-fit ``happy accident'' model has $f=0.23$, while earlier results from the physical-statistical model indicated a visual geometric albedo of 0.24. \added{This slight difference stems from the ``happy accident'' model having a shape that levels-off between the phase angles spanned by the \textit{DSCOVR} data and full phase while the physical-statistical model has a slight increase in reflectivity across this range of phase angles due to backscatter effects.}

%\,---\,that, like the earlier models, is constrained by observations, including the \textit{DSCOVR} data are near-full phase\,---\,leveling off at small phase angles while the physical models have increasing backscatter effects across as the phase angle reduces towards full phase.

%%%
%
\section{Discussion} \label{sec:discuss}
%
%%%

The discussion that follows begins with the more-narrow perspective on how the Earth planetary reflectance quantities constrained above compare to previous estimates and to those of other solar system worlds. After these details, the work above is then placed within the context of ``Earth as an exoplanet analog'' studies. Finally, areas for future improvements are discussed.

\subsection{Earth Results: Comparison to Earlier Estimates}
The visual geometric albedo of Earth arrived at in this work ($A_{\rm g}$=0.24) is markedly smaller than often-quoted values. Potentially the most-cited value is 0.367, which is stated in \textit{Allen's Astrophysical Quantities} \citep{cox2000} and derives from historical Earthshine observations in \citet{danjon1928,danjon1954}. The Danjon value required extrapolation of Earthshine-derived Earth brightness measures from about 18\textdegree~(the smallest Earth phase angle observed) to full phase. Two distinct issues lead to an over-estimation of Earth's geometric albedo. First, and has been previously analyzed \citep{qiuetal2003}, the Lunar opposition effect, and its impact on the Moon's phase curve, had not yet been well-quantified by the time of Danjon's work. As the Earth light observed in the Earthshine configuration is in the direct backscattering geometry, ignorance of the Lunar opposition effect leads to an over-estimation of Earth's reflectivity.

A second issue, not previously discussed, with the analysis towards Earth's visible geometric albedo in \citet{danjon1928,danjon1954} is an assumption that Earth's visual magnitude decreases linearly with decreasing phase angle from the smallest observed phase angle to full phase, implying an exponential increase in flux. A more-physical extrapolation could have relied on, for example, a Lambertian phase curve, and would have yielded a visual geometric albedo of 0.31. Thus, about half of the overestimation error in the apparent albedo value derived by/from \citet{danjon1928} is due to the assumed functional form of the extrapolation.

Another Earth geometric albedo estimate that appears in fact sheets is 0.434, which is even larger than the \citet{danjon1928,danjon1954} value and comes from solar system photometry work in \citet{mallamaetal2017}. This particular value\,---\,which is V-band\,---\,was determined by extrapolating optical \textit{EPOXI} photometry from 57.7\textdegree~to full phase using a sophisticated model. Unfortunately, the adopted spectral model \citep{tinettietal2006a} is known to have an error where the azimuthal angle in the underlying plane-parallel radiative transfer calculations is flipped by 180\textdegree, effectively swapping forward and backward scattering \citep{robinsonetal2011}. As cloud forward scattering is generally strong than backward scattering, this error presents as an over-estimation of Earth's full-phase brightness. Adopting the bespoke, high-fidelity \textit{EPOXI} models described in \citet{robinsonetal2011} and extrapolating to full phase yields a value for Earth's visual geometric albedo much nearer to 0.24. \added{More-recent works developing high-fidelity models of Earth's phase- and wavelength-dependent brightness arrive at broadband visual geometric albedo values of 0.26 \citep{roccettietal2025a}, in good agreement with the physically-informed value inferred here.}

A less-cited, but more-accurate, Earth visual geometric albedo value of 0.2 is provided by \citet{mallama2009}. This estimate stems from taking a reported visual spherical albedo of Earth of 0.3 \citep{palleetal2003} and dividing this by a phase integral appropriate for a Lambert sphere ($q_{\rm L}=3/2$). As the analysis presented above arrives at a visual phase integral for Earth of 1.22, the adoption of the Lambertian phase integral turns out to be an adequate approach and arrives at a more-realistic visual geometric albedo.

\subsection{Earth Results: Comparison to Other Solar System Worlds}
A visual geometric albedo of 0.24 for Earth places our planet nearer in reflectance to Mercury or Mars (visual geometric albedos of 0.14 and 0.17, respectively) than to the gas/ice giants (visual geometric albedos of 0.4--0.5) or Venus (visual geometric albedo of 0.7) \citep{mallamaetal2017,lietal2018jupiter}. Coincidentally, Earth's visual geometric albedo is quite similar to \textit{Cassini}-derived green/red values reported for Titan  that span 0.22--0.28 \citep{garciamunozetal2017}. Of course, Earth's relatively small visual geometric albedo, as compared to many other solar system planets, is primarily due to a unique feature: oceans (that cover 70\% of the planet) are very low-reflectivity when viewed in backscatter.

Earth's visual phase integral of 1.22 is comparable to reliable values reported for Jupiter at similar wavelengths \citep{lietal2018jupiter} but markedly less than phase integral values at green/red wavelengths of roughly $q=2$ reported for Titan \citep{garciamunozetal2017}. The Earth value is also somewhat smaller than that for a Lambert sphere (1.5). As shown in Figure~\ref{fig:earth_phaseint}, Earth's increased reflectivity at crescent phases more strongly emphasizes these phases (and de-emphasizes gibbous phases) in the phase integral as compared to a Lambert sphere. Titan, with its incredibly strong aerosol forward scattering seen at crescent phases, is a more-extreme example of this trade \citep{garciamunozetal2017}.

Finally, Earth's visual phase function (Figure~\ref{fig:earth_phase}) is decidedly non-Lambertian, \added{as stand-alone Earthshine observations \citep{palleetal2003} and high-fidelity models \citep[e.g.,]{tinettietal2006a,oakley&cash2009,robinsonetal2010,zuggeretal2010,roccettietal2025a} have previously attested}. The sub-Lambertian and super-Lambertian behaviors at gibbous and crescent phases, respectively, are similar to, but slightly more extreme than, non-Lambertian effects seen in Jupiter phase curves \citep{mayorgaetal2016}. However, even with aerosol forward scattering and glint, Earth's phase curve does not achieve the types of extreme crescent-phase brightness enhancements seen for Titan \citep{garciamunozetal2017,cooperetal2025}. The analysis above finds no evidence for Earth having a Moon-like opposition surge. In fact, more-recent, Earth-science-focused studies on \textit{DSCOVR} observations of Earth at phase angles approaching 2\textdegree~also do not see any opposition surge effects \citep{penttilaetal2022}.

\subsection{Earth as an Exoplanet Analog}
Analog studies that treat Earth as an exoplanet aim to set aside prior knowledge and understand observations of our planet using exoplanetary science tools and techniques. It is often challenging, though, to truly set aside all prior knowledge. For the wavelength- and phase-dependent models explored here, the approach was not completely blind. The models assume the presence of a surface and pure-scattering clouds, adopt an Earth-like stratospheric column abundance of ozone, and use an Earth-like column Rayleigh scattering optical depth. These assumptions are justifiable as retrievals on simulated reflected-light direct imaging observations of Earth (as would be obtained by, e.g., HWO) do not struggle to detect a surface, clouds, Rayleigh scattering, and ozone \citep{fengetal2018,damiano&hu2021,latoufetal2023,gomezbarrientosetal2023,salvadoretal2024,tokadjianetal2024,ulsesetal2025}. 

The log-Bayes factors for models shown in Table~\ref{tab:models} demonstrate that vertically optically thin aerosol scattering, with or without ocean glint, is required to reproduce Earth's enhanced apparent albedo at crescent phases. Of the best-performing models (Models 02, 05, and 07), those that include an ocean glint treatment (Models 05 and 07) represent a ``true positive'' detection of surface habitability. Model 02, which does not include an ocean glint contribution, would represent a false negative for the detection of surface habitability as it can match the aerosol and glint contributions in Earth's phase curve using only aerosol forward scattering. \added{That these models all have comparable log-Bayes factors thus indicates that the phase curve analysis does not yield a surface habitability detection.}

The false negative for surface habitability seen in Model 02 and produced by aerosol forward scattering does not spell doom for surface habitability detections from phase curves; the visual range is known to be a poor band for glint detection as Rayleigh scattering obscures surface effects \citep{robinsonetal2010,zuggeretal2011a,vaughanetal2023}. Phase curves observed at redder-optical or near-infrared continuum wavelengths \added{(where Earthshine observations are rarely taken)} would have markedly better ocean glint sensitivity. Observations at non-continuum wavelengths could also help to disentangle aerosol effects from glint\,---\,recent work on near-infrared Titan phase curves shows that haze forward scattering (as indicated by the crescent-to-gibbous brightness ratio) is enhanced in gas absorption features, which is the opposite trend expected for ocean glint \citep{cooperetal2025}. \added{Potentially more powerfully, polarimetric phase curves, especially at red-optical wavelengths, have been shown to be an important avenue for constraining surface habitability \citep{roccettietal2025c}.}

Moving beyond glint detections, the retrieval methods developed here for application to phase curves demonstrate that important information can be inferred from phase functions and that this information complements that gained from spectral observations. Importantly, inferences from phase curves can be accomplished with photometric observations that demand less exposure time than spectroscopic observations. The posteriors shown in Appendix~\ref{sec:corners} include inferences of the global coverage of optically thick clouds and the asymmetry parameter for backscattering from these clouds (which is constrained by Earth's slightly-increasing apparent albedo when the illumination is increased from quadrature to full phase). The coverage of thick clouds in Models 02, 05, and 07 is 10--20\%, which is somewhat smaller than the 30--60\% value sometimes quoted for thick clouds \citep{wood2012}. Potential explanations for this difference include, first, that the vertically optically thin aerosol component in the adopted models is likely also capturing some aspects of water cloud scattering and, second, the phase curve observations used here are more equator-dominated, which is a region of Earth that is known to be less cloudy \citep{katoetal2019}. \added{Finally, a different fitted cloud fraction could be preferred if more-sophisticated, three-dimensional clouds treatments were adopted \citep{roccettietal2025a,roccettietal2025b,roccettietal2025c}.} Other parameters for the cloud single-scattering phase function are not well-constrained.

It is intriguing that Models 02, 05, and 07 find ``thin aerosol'' vertical optical depths of 0.1--0.3 with only weak forward scattering. This optical depth scale is in line with typical values for cirrus clouds \citep{heidingeretal2015} and for standard clearsky aerosol optical depth measurements \citep{feietal2019}. The fitting preference for more-isotropic scattering from these aerosols may stem from the assumption of pure scattering, where allowing the aerosols to have a non-unity single scattering albedo could maintain the proper apparent albedo contribution at high phase angles with a larger asymmetry parameter. As a related aside, large volcanic eruptions (e.g., the 1991 eruption of Mount Pinatubo) can increase stratospheric aerosol optical depths by more than 0.1 \citep{vernieretal2011}, implying phase curve monitoring for Earth-like exoplanets could reveal similar eruptive events.

\added{For Model 07, which is the most Earth-like model in its treatments, the gray Lambertian surface albedo (attributed to land) is only loosely constrained to be below 0.4 (at 84\% confidence), which is consistent with typical land and vegetation surface flux albedo values at these wavelengths \citep[][, their Figure~1]{ulsesetal2025}. Surface wind speeds are not well constrained, but are limited by the priors adopted for the underlying physical model. Lastly, the land-ocean ratio for Earth is known to be 0.4, which coincides with a weak peak in the inferred value for $r_{\rm l/o}$ for Model 07.}

Finally, all of Models 02, 05, and 07 find a breakpoint phase angle\,---\,where the phase curve variability transitions from constant with phase angle to a power-law\,---\,of roughly 100\textdegree. The increasing phase curve variability with increasing phase angle through crescent phases is driven by the ability of the illuminated sliver of Earth's disk to be dominated by increasingly smaller structures (e.g., storms). Thus, the breakpoint angle likely identifies the spatial scale at which the size of the illuminated portion of the disk becomes comparable to the largest-scale weather structures. The breakpoint angle, however, may be sensitive to the adopted statistical model for variability, and future work could investigate if a Gaussian variability model is appropriate for some/all Earth phases. \added{Additionally, as the variability model is, at some level, sensitive to the scale of weather patterns on Earth, future work would need to address whether such a breakpoint argument applies for other worlds.}

\subsection{Considerations and Areas for Improvement}
Above all else, the results above\,---\,especially the recognition of an aerosol scattering false negative for surface habitability\,---\,demand more observations of Earth's phase curve, especially at redder and/or near-infrared wavelengths. The improved surface sensitivity of these phase curves could help to better disentangle glint effects from aerosol scattering. Additionally, jointly fitting two (or more) lightcurves spanning different spectral ranges could lead to more-confident glint detections.

A complication avoided in the analyses above is that the radius of a directly-imaged exoplanet is likely to be only loosely constrained by an observed planet-to-star flux ratio spectrum \citep{fengetal2018,salvadoretal2024,damianoetal2025}.  This uncertainty would propagate into any derived phase curve, as the inferred phase-dependent brightness, $A_{\rm g} \Phi(\alpha)$, depends on the planet-to-star flux ratio and the square of the planetary radius (and the square of the orbital distance). The models used here predict the true (i.e., unscaled) value of the phase-dependent brightness, so future applications to exoplanet phase curve data \added{from direct imaging} would either need to propagate through a radius uncertainty or explore fits with scaled models. The latter approach abandons information gained from the true value of the phase-dependent brightness and relies only on phase curve shape.

Direct imaging of exoplanets in reflected light can only reveal limited portions of a world's phase curve, depending on the orbit and the inner working angle of the imaging system, and are expected to be both sparser and at lower signal-to-noise than the observations explored here. Thus, future efforts should explore how limiting access to regions of Earth's phase curve, sparser sampling, \added{added polarimetry}, and/or decreased signal-to-noise ratios all impact the derived inferences. Such studies could also explore joint spectral and phase curve retrievals, although the computational expense of accurately modeling spectra and/or photometry at a number of phase angles may prove limiting. Clearly we are only beginning to understand what environmental information could be gleaned from reflected-light phase curves of directly imaged exoplanets.

%%%
%
\section{Conclusions} \label{sec:conc}
%
%%%

Analyses of reflected-light phase curves can reveal fundamental planetary properties, are complementary to spectral characterization approaches, and will be a growing area of research given planned observations with the \textit{Roman} Coronagraph Instrument \citep{poberezhskiyetal2022} and HWO \citep{feinbergetal2024}. For Earth, approaches to understanding its phase-dependent brightness are especially significant as these observations are generally challenging to achieve. While the efforts in this manuscript are focused on inferring key reflectance-related quantities for Earth (where some earlier results are in error), the methods developed can be expanded to phase curve analyses for worlds beyond Earth. Key products, findings, and tools from this work include:

\begin{itemize}
    \item A dataset for Earth's phase-dependent visual \added{(0.4--0.7\,$\upmu$m)} brightness\,---\,expressed as apparent albedo\,---\,is curated and spans 5\textdegree~to 144\textdegree~in phase angle (Figure~\ref{fig:earth_data}).
    \item A new inverse method is developed that represents phase-dependent brightness through a physical-statistical model and that includes scattered light contributions from optically thick clouds, vertically optically thin aerosols, ocean glint, Rayleigh scattering, gas absorption, and Lambertian surface reflectance (Section~\ref{sec:model}).
    \item Fitting the physical-statistical model to the curated Earth phase curve dataset provides constraints on Earth's visual \added{(0.4--0.7\,$\upmu$m)} geometric albedo ($0.242^{+0.005}_{-0.004}$), visual spherical albedo ($0.294^{+0.002}_{-0.002}$), visual phase integral ($1.22^{+0.02}_{-0.03}$), and geometric albedo in \added{B (0.4--0.5\,$\upmu$m), V (0.5--0.6\,$\upmu$m), and R (0.6--0.7\,$\upmu$m) bands} are ($0.277^{+0.005}_{-0.004}$, $0.226^{+0.004}_{-0.004}$, and $0.221^{+0.004}_{-0.004}$, respectively).
    \item Analysis of a best-fitting physical-statistical model reveals the relative contributions of different scattering processes to Earth's phase-dependent brightness (Figure~\ref{fig:process_swaths}). This analysis also quantifies the non-Lambertian nature of Earth's planetary phase function (Figure~\ref{fig:earth_phase}).
    \item Model selection reveals that thin aerosol forward scattering can yield a false negative for detecting surface habitability through ocean glint effects. Observations and analysis of phase curves at redder-optical and/or near-infrared wavelengths\,---\,that have improved surface sensitivity\,---\,would help to understand how to avoid this false negative.
    \item \added{An accurate reproduction of Earth's visual phase curve comes from representing it with a normalized Henyey-Greenstein phase function (Equation~\ref{eqn:henyey}) and is achieved using the two-parameter model in Equation~\ref{eqn:happyacc} with $f=0.23$ and $g=-0.33$.}
\end{itemize}

%%%
%
\acknowledgements{TDR gratefully acknowledges support from NASA's Habitable Worlds Program (No.~80NSSC20K0226), NASA's Exoplanet Research Program (No.~80NSSC25K7149), and the Nexus for Exoplanet System Science Virtual Planetary Laboratory (No.~80NSSC23K1398). TDR thanks J.~Lustig-Yaeger for key discussions about nested sampling methods, E.~Agol for a conversation about statistical models, V.~Meadows for comments on an early concept for this manuscript, J.~Pepper for inviting a presentation on ``Earth as an exoplanet'' that precipitated some thinking developed here, and all SAG~26 participants for their dedication to improving radiative transfer tools. E.~Pall{\'e} and S.~Fan are thanked for sharing Earthshine and \textit{DSCOVR} data, respectively. Finally, G.~Roccetti and two anonymous reviewers are thanked for providing friendly, constructive reviews of the originally-submitted version of this manuscript, and A.~Zelakiewicz is thanked for spotting a missing minus sign in an earlier expression of the Henyey-Greenstein phase function.}
%
%%%

%%%
%
\appendix

\section{Review of Planetary Photometry Theory}
\label{sec:photom_review}

The following review spans reflectance properties beginning at planetary surfaces and expanding to disk-averaged quantities. A motivation behind this review is to provide, in a single location, a description of the parameters and processes required for studying the many facets of planetary photometry. A secondary motivation is that, unfortunately, widely-used textbooks can be in error when it comes to deriving and defining the quantities detailed below. For convenience, Table~\ref{tab:vars} summarizes all of the parameters that appear in this section.

Explicit dependence on wavelength is omitted here for conciseness for all quantities except when discussing the Bond albedo. Radiant intensity (with units of energy per unit area per unit time per unit solid angle) and flux density (simply flux hereafter; with units of energy per unit area per unit time) can either be specific (i.e., per unit wavelength, frequency, or wavenumber) or spectrally integrated. In the latter case, the spectral integration must occur before any intensity or flux terms are divided to yield a reflectance-related quantity.

The surface bidirectional reflectance distribution function (BRDF) codifies how a surface reflects incident radiation into upwelling intensity. Formally, the BRDF can vary over the entire planetary surface and in time. In practice, though, it is useful to apply the BRDF assuming that a patch of planetary surface is planar and homogeneous. Following \citet[][their Section~5.2.4]{thomas&stamnes1999}, the upwelling intensity emerging from a surface, $I\left(\mu>0,\phi\right)$ (where $\mu$ is the cosine of the zenith angle and $\phi$ is the azimuth angle), resulting from the direction-dependent reflectance of a downwelling intensity field, $I\left(\mu^{\prime}<0,\phi^{\prime}\right)$, is expressed as,
\begin{equation}
    I\left(\mu>0,\phi\right) = \int_{0}^{2\pi} \int_{-1}^{0} \rho\left(\mu^{\prime},\phi^{\prime},\mu,\phi\right) \cdot I\left(\mu^{\prime},\phi^{\prime}\right) \mu^{\prime}d\mu^{\prime}d\phi^{\prime} \ ,
    \label{eqn:brdf}
\end{equation}
where the integral is over all downwelling directions, $\rho$ is the BRDF, and the collection of terms following the BRDF represent the energy flux incident on the surface. The surface flux albedo, commonly $A$ in textbooks, is the ratio of the emergent flux from the surface to the incident flux, or,
\begin{equation}
    A \equiv \frac{ \int_{0}^{2\pi} \int_{0}^{1} I\left(\mu,\phi\right) \mu d\mu d\phi }{ \int_{0}^{2\pi} \int_{-1}^{0} I\left(\mu^{\prime},\phi^{\prime}\right) \mu^{\prime}d\mu^{\prime}d\phi^{\prime} } \ .
    \label{eqn:fluxalb}
\end{equation}
The numerator on the right hand side of this expression can be expressed in terms of the BRDF using the preceding expression. Given the BRDF dependence in the numerator, the flux albedo for a collimated direct beam of solar flux need not be the same as the flux albedo for diffuse flux. Nevertheless, the surface flux albedo is often taken to be independent of the flux source.

It is instructive to consider the scenario where the surface is Lambertian, so that the BRDF is independent of the incident and emergent geometry. Inserting,
\begin{equation}
    \rho\left(\mu^{\prime},\phi^{\prime},\mu,\phi\right) = \rho_{\rm L} \ ,
\end{equation}
into Equation~\ref{eqn:brdf} yields,
\begin{equation}
    I\left(\mu>0,\phi\right) = \rho_{\rm L} \int_{0}^{2\pi} \int_{-1}^{0} I\left(\mu^{\prime},\phi^{\prime}\right) \mu^{\prime}d\mu^{\prime}d\phi^{\prime} = \rho_{\rm L} F^{-} \ ,
\end{equation}
where $F^{-}$ is the downwelling flux incident on the surface. For a collimated direct beam of solar flux at a solar zenith angle cosine of $-\mu_{\rm s}$, the downwelling flux at a time-dependent planetary orbital distance, $r$, is $\mu_{\rm s} F_{\rm s}(r)$ and the previous expression yields Lambert's cosine law \citep{lambert1760},
\begin{equation}
    I\left(\mu>0,\phi\right) = \rho_{\rm L} \mu_{\rm s} F_{\rm s}\!\left( r \right) \propto \cos \theta_{\rm s} F_{\rm s}\!\left( r \right) \ ,
\end{equation}
where $\theta_{\rm s}$ is the solar zenith angle. Adopting the Lambertian surface assumption into Equation~\ref{eqn:fluxalb} yields a Lambert flux albedo,
\begin{equation}
    A_{\rm L} = \pi \rho_{\rm L} \ .
\end{equation}

Subsequent reflectance-related quantities discussed here are generally applied to planetary disks or spheres and, thus, require integrals of spatially- and temporally-varying quantities over solid angle. Furthermore, while the preceding materials involved intensities and fluxes at a planar surface, the planetary-scale materials that follow apply to an external observer and are, thus, top of atmosphere. Only if atmospheric effects are omitted or negligible do the preceding BRDF materials apply to an external observer.

Rather than immediately imposing a latitude/longitude coordinate system, generality can be maintained by defining unit vectors pointing radially outward from the planet at the sub-observer and sub-solar points (${\bf \hat{o}}$ and ${\bf \hat{s}}$, respectively) and ${\bf \hat{n}}$ as a similar unit vector for any arbitrary latitude/longitude point on the planet. The planetary phase angle (i.e., the star-planet-observer angle) is then given by,
\begin{equation}
    \cos \alpha = \cos \alpha\!\left( {\bf \hat{o}}(t), {\bf \hat{s}}(t) \right) = {\bf \hat{o}} \cdot {\bf \hat{s}} \ ,
\end{equation}
where the time ($t$) dependence in the sub-observer and sub-solar locations is omitted hereafter for conciseness.

\begin{figure}
    \centering
    \includegraphics[scale=0.5,trim=200mm 20mm 200mm 20mm]{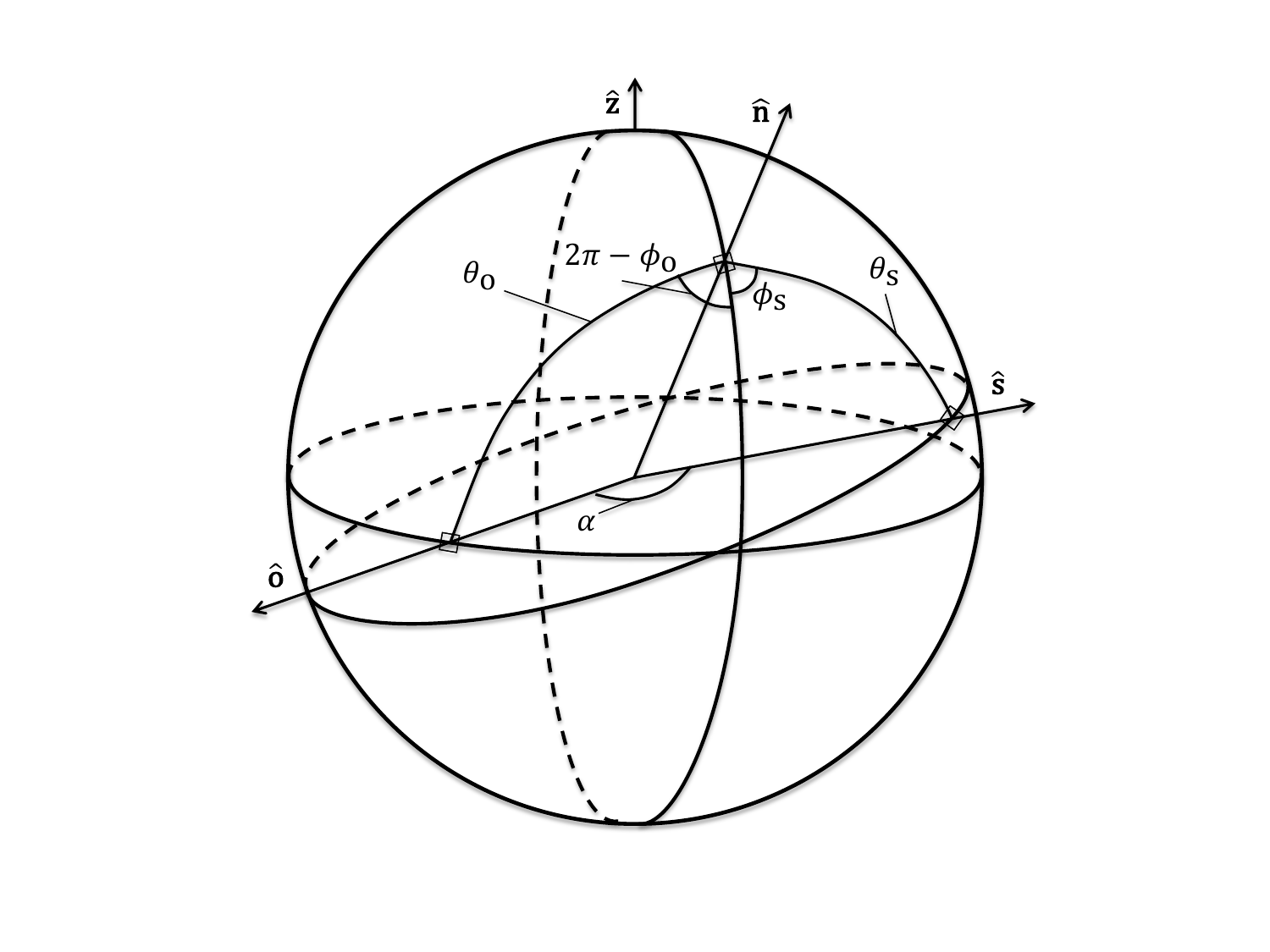}
    \caption{Vector quantities and angles relevant to planetary disk integration represented on a unit sphere. Unit vectors ${\bf \hat{o}}$ and ${\bf \hat{s}}$ indicate the direction of the observer and sun, respectively. Angles $\theta_{\rm o}$ and $\theta_{\rm o}$ are zenith angles for the observer and sun, respectively, in a polar coordinate system at an arbitrary location indicated by the unit vector ${\bf \hat{n}}$. Angles $\phi_{\rm o}$ and $\phi_{\rm s}$ are azimuth angles for the observer and sun, respectively, in this same polar coordinate system.}
    \label{fig:geom_vec}
\end{figure}

If an observer measures the flux ($F$) from a planetary disk at known observer-planet distance ($d$) and phase angle, the planetary disk-averaged intensity is then,
\begin{equation}
    \bar{I}\left (\alpha \right) \equiv \frac{ F \left( d,\alpha \right) }{\pi (R_{\rm p}/d)^2 } \ ,
    \label{eqn:avgint}
\end{equation}
where $R_{\rm p}$ is the planetary radius. In solar system science, where targets are often resolved, a commonly used measure is ``$I/F$,'' which is the ratio of the intensity recorded (e.g., by a CCD pixel) to the normal-incidence solar flux at the planetary orbit. From such resolved imaging, the geometric albedo can be equivalently defined via an integral over the apparent solid angle of the disk, $\Omega_{\rm p}(d)$, with,
\begin{equation}
    \bar{I}\left (\alpha \right) = \frac{1}{\pi (R_{\rm p}/d)^2} \int_{\Omega_{\rm p}(d)} I/F \cdot F_{\rm s}(r) d\omega \ ,
\end{equation}
where, as before, $F_{\rm s}(r)$ is the normal-incidence solar flux at the planetary orbital distance. In practice, this quantity is obtained by a summation over pixels encompassing the planetary disk \citep{mayorgaetal2016}. Spatially-resolved models generally simulate the top of atmosphere intensity for a location on the disk in the direction of an observer given the location of the sub-solar point, represented here as $I( {\bf \hat{n}},{\bf \hat{o}},{\bf \hat{s}} )$ (that, when divided by $\mu_{\rm s} F_{\rm s}$, yields the ``reflection coefficient) \citep[][]{dlugach&yanovitskij1974,hengetal2021} . The projected disk-averaged intensity is then given by,
\begin{equation}
    \bar{I}\left (\alpha \right) = \frac{1}{\pi} \int_{2\pi} I\left({\bf \hat{n}},{\bf \hat{o}},{\bf \hat{s}} \right) \left( {\bf \hat{n}} \cdot {\bf \hat{o}}\right) d\omega ,
    \label{eqn:avgint_integral}
\end{equation}
where the integral is over a hemisphere of the globe (i.e., latitude and longitude are the variables of integration), ${\bf \hat{n}}$ corresponds to the location on the globe for the infinitesimal solid angle unit $d\omega$, and the dot product ensures that equal-area solid angle units near the limb are weighted less due to projection effects. In the local polar coordinate system at ${\bf \hat{n}}$, the observer and solar zenith angle cosines and azimuth angles are,
\begin{equation}
    \mu_{\rm o} = \cos \theta_{\rm o} = {\bf \hat{n}} \cdot {\bf \hat{o}} \ ,
\end{equation}
\begin{equation}
    \mu_{\rm s} = \cos \theta_{\rm s} = {\bf \hat{n}} \cdot {\bf \hat{s}} \ ,
\end{equation}
\begin{equation}
    \tan \phi_{\rm o} = \frac{{\bf \hat{n}} \cdot \left( {\bf \hat{o}} \times {\bf \hat{z}} \right)}{{\bf \hat{o}} \cdot {\bf \hat{z}} - \left( {\bf \hat{n}} \cdot {\bf \hat{o}} \right)\left( {\bf \hat{n}} \cdot {\bf \hat{z}} \right)} \ ,
\end{equation}
and
\begin{equation}
    \tan \phi_{\rm s} = \frac{{\bf \hat{n}} \cdot \left( {\bf \hat{s}} \times {\bf \hat{z}} \right)}{{\bf \hat{s}} \cdot {\bf \hat{z}} - \left( {\bf \hat{n}} \cdot {\bf \hat{s}} \right)\left( {\bf \hat{n}} \cdot {\bf \hat{z}} \right)} \ ,
\end{equation}
where ${\bf \hat{z}}$ is a unit vector along the axis for which the polar angle is measured on the globe (e.g., directed at the northern pole) and the azimuth angle expressions lend themselves to the two-argument arctangent function. Only locations that satisfy ${\bf \hat{n}} \cdot {\bf \hat{o}} > 0 $ are observable and locations that satisfy ${\bf \hat{n}} \cdot {\bf \hat{s}} > 0 $ are receiving solar flux. Figure~\ref{fig:geom_vec} helps to visualize these vector quantities and angles on a unit sphere.

The disk-averaged intensity terms can be used to define the geometric albedo ($A_{\rm g}$), which indicates the reflectivity of a world observed at full phase ($\alpha=0$; i.e., where the sub-solar and sub-observer locations are identical). This albedo is formally defined as the ratio of the planetary flux received at full phase to the flux received from a flat, perfectly-reflecting Lambert disk of identical size and observed at the same distance. This latter quantity is simply $F_{\rm s}(r)(R_{\rm p}/d)^2$. Thus, in terms of intensities, the geometric albedo is,
\begin{equation}
    A_{\rm g} \equiv \frac{ \bar{I}\left (\alpha=0 \right) }{ F_{\rm s}\left( r \right)/\pi } \ ,
    \label{eqn:geomalb}
\end{equation}
where the denominator is arranged to equal the intensity of the flat, perfectly-reflecting Lambert disk. For the geometric albedo definition, and for all subsequent albedos, it is assumed that the planetary flux is only due to reflected light. From an observational perspective, this either means that the planetary thermal flux is well-separated in wavelength or that the thermal emission has been accounted for using a model. Separating scattered sunlight from thermal emission is straightforward for models as these sources combine linearly in the radiative transfer equation. Most generally, the geometric albedo can vary in time due to, e.g., planetary rotation, weather on the observable hemisphere, and/or obliquity-related effects that cause the sub-solar location to vary over an orbit.

Variations in the brightness of a planetary disk with phase are usually represented using the phase function, 
\begin{equation}
    \Phi\left( \alpha \right) \equiv \frac{\bar{I}\left( \alpha \right)}{\bar{I}\left( 0 \right)} \ .
    \label{eqn:phasefunc}
\end{equation}
In general, the phase-dependent brightness of a planet is not simply a function of a single parameter like the phase angle \citep[see discussion of the ``generalized'' phase law in][]{lesteretal1979}. A target can be observed twice with identical orientations of the sun and observer and, yet, have different disk-averaged intensities for the two epochs due to, e.g., weather. A convenient example comes from observations of Earth from NASA's \textit{EPOXI} mission \citep{livengoodetal2011}, where data acquired at pole-on geometries produced brighter disk-averaged intensities than for similar-phase equator-on observations \citep{cowanetal2009}. Combining Equations~\ref{eqn:geomalb} and \ref{eqn:phasefunc} with the definition of the disk-averaged intensity (Equation~\ref{eqn:avgint}) yields the planet-to-star flux ratio often used in discussions of exoplanet direct imaging,
\begin{equation}
    \frac{F_{\rm p}}{F_{\rm s}} = \frac{F\left( d,\alpha \right)}{F_{\rm s}\left( d \right)} = A_{\rm g} \Phi\left( \alpha \right) \left( \frac{R_{\rm p}}{r} \right)^2 \ .
    \label{eqn:FpFs}
\end{equation}

The spherical albedo is a wavelength-dependent quantity given by the ratio of the power reflected by a planet to the solar power incident on the planet. At a given location on the globe, the upwelling flux in reflected light can be obtained by integrating the emergent intensity over the local upper hemisphere,
\begin{equation}
    F^{+}\left( {\bf \hat{n}},{\bf \hat{s}} \right) = \int_{0}^{2\pi} \int_{0}^{1} I\left(\mu,\phi,{\bf \hat{n}},{\bf \hat{s}} \right) \mu d\mu d\phi \ .
\end{equation}
Integrating the location-dependent upwelling flux over the globe and dividing by the solar power intercepted by the planet gives the spherical albedo,
\begin{equation}
    A_{\rm s} \equiv \frac{1}{\pi R_{\rm p}^2 F_{\rm s}\left( r \right)} \int_{2\pi} F^{+}\left( {\bf \hat{n}},{\bf \hat{s}} \right) R_{\rm p}^2 d\omega  = \frac{1}{\pi F_{\rm s}\left( r \right)} \int_{2\pi} F^{+}\left( {\bf \hat{n}},{\bf \hat{s}} \right) d\omega \ ,
    \label{eqn:sphere_alb}
\end{equation}
where the integral is over the illuminated hemisphere.

Separately integrating the numerator and denominator of the spherical albedo definition over all wavelengths\,---\,to obtain the bolometric power reflected by the planet and the bolometric solar power intercepted by the planet\,---\,yields the Bond albedo ($A_{\rm B}$). Equivalently, the Bond albedo can be calculated by weighting the wavelength-dependent spherical albedo by the incident specific solar flux and integrating over wavelength, 
\begin{equation}
    A_{\rm B} \equiv \frac{\int_{0}^{\infty} A_{\rm s}\left( \lambda \right) F_{{\rm s},\lambda} d\lambda}{\int_{0}^{\infty} F_{\rm s,\lambda} d\lambda} \ ,
    \label{eqn:bond_alb}
\end{equation}
where the wavelength dependence in the spherical albedo and the specific nature of the solar flux have been made explicit for clarity. The Bond albedo can also be inferred if the planetary effective temperature is known, provided the world has no substantial internal heat source and assuming the planetary emitted power to be in balance with the absorbed power.

In practice it is prohibitively challenging to measure the reflected light radiation field emerging from an entire planetary globe. A simplification can be made where the intensity emerging from a location on the globe is isotropic and that the disk-averaged intensity is only a function of the phase angle (which is an assumption that works best for worlds with more-homogeneous atmospheres and surfaces). The wavelength-dependent power reflected by the planet can then be reduced to an integral over phase angle (or, from an observational perspective, an integral over disk-averaged intensities measured over a complete set of phase angles), with,
\begin{equation}
    A_{\rm s} = \frac{2}{F_{\rm s}\!\left( r \right)/\pi} \int_{0}^{\pi} \bar{I}\left( \alpha \right) \sin \alpha d\alpha = A_{\rm g} \cdot 2\int_{0}^{\pi} \Phi \left( \alpha \right) \sin \alpha d\alpha = A_{\rm g} \cdot q \ ,
    \label{eqn:sphere_alb_simp}
\end{equation}
where the phase integral has been introduced as,
\begin{equation}
    q \equiv 2\int_{0}^{\pi} \Phi \left( \alpha \right) \sin \alpha d\alpha \ .
    \label{eqn:phaseint}
\end{equation}
From a planetary exploration and radiative balance perspective, the integrand of the phase integral, $\Phi(\alpha)\sin \alpha$, is a useful quantity for understanding which phases have the strongest contribution to the spherical albedo \citep{garciamunozetal2017}.

\begin{figure}
    \centering
    \includegraphics[scale=0.5,trim=200mm 20mm 200mm 20mm]{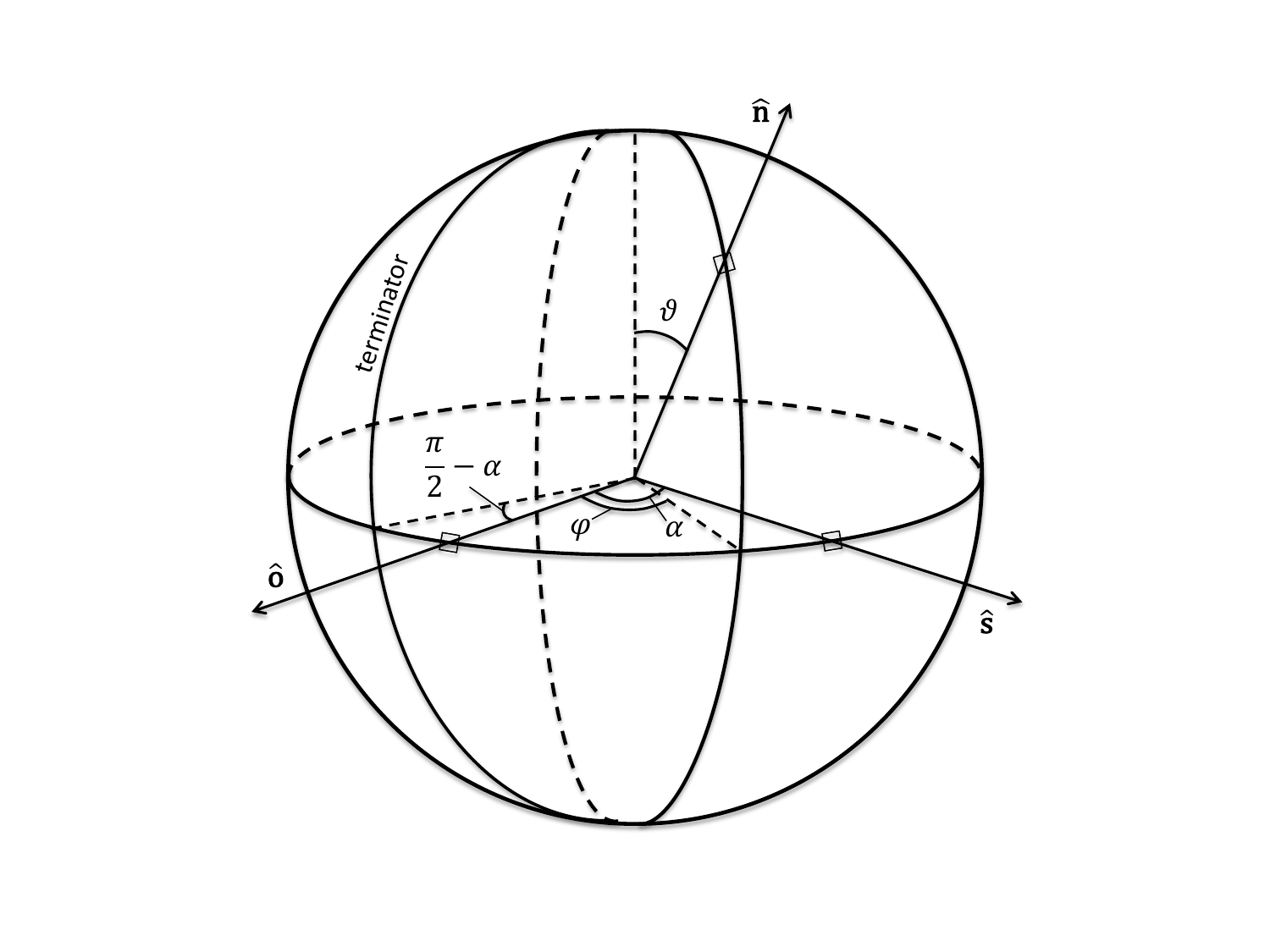}
    \caption{An alternative coordinate system relevant to planetary disk integration \citep[after][]{sobolev1975} where the equator is defined by the great arc that connects the sub-observer and sub-solar locations. A polar angle, $\upvartheta$, is measured from a pole given this ``intensity equator,'' and an azimuthal angle, $\upvarphi$, is measured in the equatorial plane relative to the sub-observer location.}
    \label{fig:geom_sob}
\end{figure}

Finally, the phase-dependent disk-averaged intensity of a so-called Lambert sphere\,---\,which assumes no atmospheric effects and a uniform Lambertian surface over the entire globe\,---\,can provide a useful scaling for measurements of planetary disk-averaged intensities. Figure~\ref{fig:geom_sob}, after \citet{sobolev1975}, provides a convenient set of coordinates where the sub-observer and sub-solar locations define an ``intensity equator,'' the polar angle ($\upvartheta$; akin to latitude) is measured relative to a pole given this equator, and the azimuthal angle ($\upvarphi$; akin to longitude) is measured in the equatorial plane relative to the sub-observer location. In these coordinates, the solar and observer zenith angle cosines at an arbitrary location, ${\bf \hat{n}}$, are then,
\begin{equation}
    \mu_{\rm s} = \sin \upvartheta \cos\left( \alpha - \upvarphi \right) \ ,
\end{equation}
and
\begin{equation}
    \mu_{\rm o} = \sin \upvartheta \cos\left( \upvarphi \right) \ ,
\end{equation}
so that the phase angle-dependent disk-averaged intensity of a Lambert sphere with surface flux albedo, $A_{\rm L}$, is (via Equation~\ref{eqn:avgint_integral}),
\begin{equation}
    \bar{I}_{\rm L}\left( \alpha, A_{\rm L} \right) = \rho_{\rm L} \frac{F_{\rm s}\!\left( r \right)}{\pi} \int_0^\pi \int_{\alpha-\pi/2}^{\pi/2} \cos\left( \alpha - \upvarphi \right) \cos \upvarphi \sin^3 \upvartheta d\upvarphi d\upvartheta =  \frac{2}{3} A_{\rm L} \frac{F_{\rm s}\!\left( r \right)}{\pi} \cdot \frac{ \sin \alpha + \left( \pi - \alpha \right) \cos \alpha}{\pi}   \ ,
\end{equation}
so that the geometric albedo of a Lambert sphere is,
\begin{equation}
    A_{\rm g,L} = \frac{2}{3} A_{\rm L} \ ,
\end{equation}
and the Lambert phase function \citep{russell1916} is,
\begin{equation}
    \Phi_{\rm L} \left( \alpha \right) = \frac{ \sin \alpha + \left( \pi - \alpha \right) \cos \alpha}{\pi} \ .
    \label{eqn:lambert}
\end{equation}
Given the Lambert phase function and the definition of the phase integral (Equation~\ref{eqn:phaseint}), the phase integral for a Lambert sphere is straightforwardly found to be $q_{\rm L}=3/2$ and the spherical albedo of a Lambert sphere is identical to the surface flux albedo (i.e., $A_{\rm s, L} = A_{\rm L}$). The apparent albedo, $A_{\rm app}$, can then be defined as the Lambert sphere surface flux albedo required to reproduce a measured planetary phase-dependent disk-averaged intensity. Thus, the apparent albedo satisfies,
\begin{equation}
    \bar{I}_{\rm L}\left( \alpha, A_{app} \right) = \bar{I}\left( \alpha \right) \ ,
\end{equation}
or
\begin{equation}
    A_{\rm app} \equiv \frac{3}{2} \frac{1}{\Phi_{\rm L} \left( \alpha \right)} \frac{ \pi \bar{I}\left( \alpha \right)}{ F_{\rm s}\!\left( r \right) } .
    \label{eqn:Aapp}
\end{equation}
The apparent albedo can be wavelength dependent, through the dependence of the planetary disk-averaged intensity on wavelength, and deviations from a constant apparent albedo with phase angle can indicate non-Lambertian scattering from the planet.

%% symbols
\begin{table}[ht]
  \centering
  {\bf Planetary Photometry Symbol Usage} \\
  \vspace{2mm}
  \begin{tabular}{c l}
    \hline
    \hline
    Symbol &  Description \\ \hline
   $A$              &   surface flux albedo \\
   $A_{\rm app}$    &   apparent albedo \\
   $A_{\rm B}$      &   Bond albedo \\
   $A_{\rm g}$      &   geometric albedo \\
   $A_{\rm s}$      &   spherical albedo \\
   $\alpha$         &   star-planet-observer (phase) angle \\
   $d$              &   distance (e.g., observer-planet) \\
   $F^{+/-}$        &   upwelling/downwelling atmospheric flux \\
   $F_{\rm s}\!(r)$ &   solar (or stellar; normal-incidence) flux at distance $r$ \\
   $F_{\rm p}\!(d)$ &   planetary flux at distance $d$ \\
   $I$              &   intensity or specific intensity \\
   $\bar{I}$        &   planetary disk-averaged intensity \\
   $I/F$            &   ratio of emergent intensity to $F_{\rm s}\!(r)$ \\
   $\mu$            &   zenith angle cosine \\
   $\mu_{\rm o}$    &   observer zenith angle cosine \\
   $\mu_{\rm s}$    &   solar (or stellar) zenith angle cosine \\
   ${\bf \hat{o}}$  &   unit vector in direction of observer \\
   $\Omega_{\rm p}$ &   apparent solid angle of planetary disk \\
   $\Phi$           &   phase function \\
   $\phi$           &   azimuthal angle \\
   $\phi_{\rm o}$   &   observer azimuthal angle \\
   $\phi_{\rm s}$   &   solar (or stellar) azimuthal angle \\
   $\upvarphi$      &   planetary globe azimuthal angle \\
   $q$              &   phase integral \\
   ${\bf \hat{s}}$  &   unit vector in direction of sun/star \\
   $\rho$           &   bi-directional reflectance distribution function (BRDF) \\
   $R_{\rm p}$      &   planetary radius \\
   $r$              &   planetary orbital distance \\
   $\rho_{\rm L}$   &   Lambertian (isotropic) BRDF \\
   $t$              &   time \\
   $\theta$         &   zenith angle \\
   $\theta_{\rm o}$ &   observer zenith angle \\
   $\theta_{\rm s}$ &   solar (or stellar) zenith angle \\
   $\upvartheta$    &   planetary globe polar angle \\
  \hline
  \end{tabular}
\caption{Relevant parameters for development of planetary photometric theory.}
\label{tab:vars}
\end{table}
\section{Corner Plots}
\label{sec:corners}

\begin{figure}
    \centering
    \includegraphics[scale=0.25,trim=0mm 0mm 0mm 0mm]{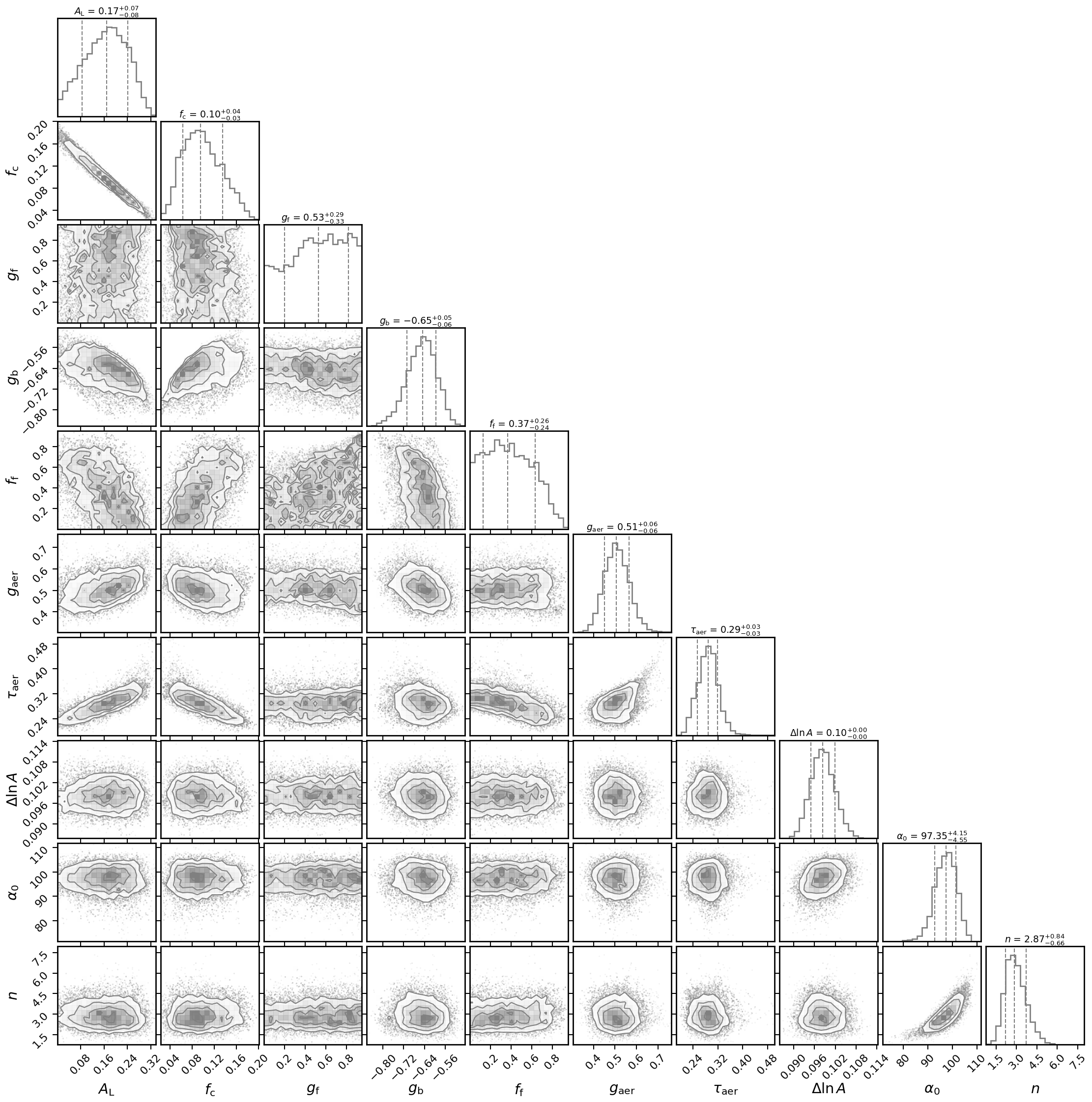}
    \caption{Corner plot for Model 02.}
    \label{fig:corner02}
\end{figure}

\begin{figure}
    \centering
    \includegraphics[scale=0.25,trim=0mm 0mm 0mm 0mm]{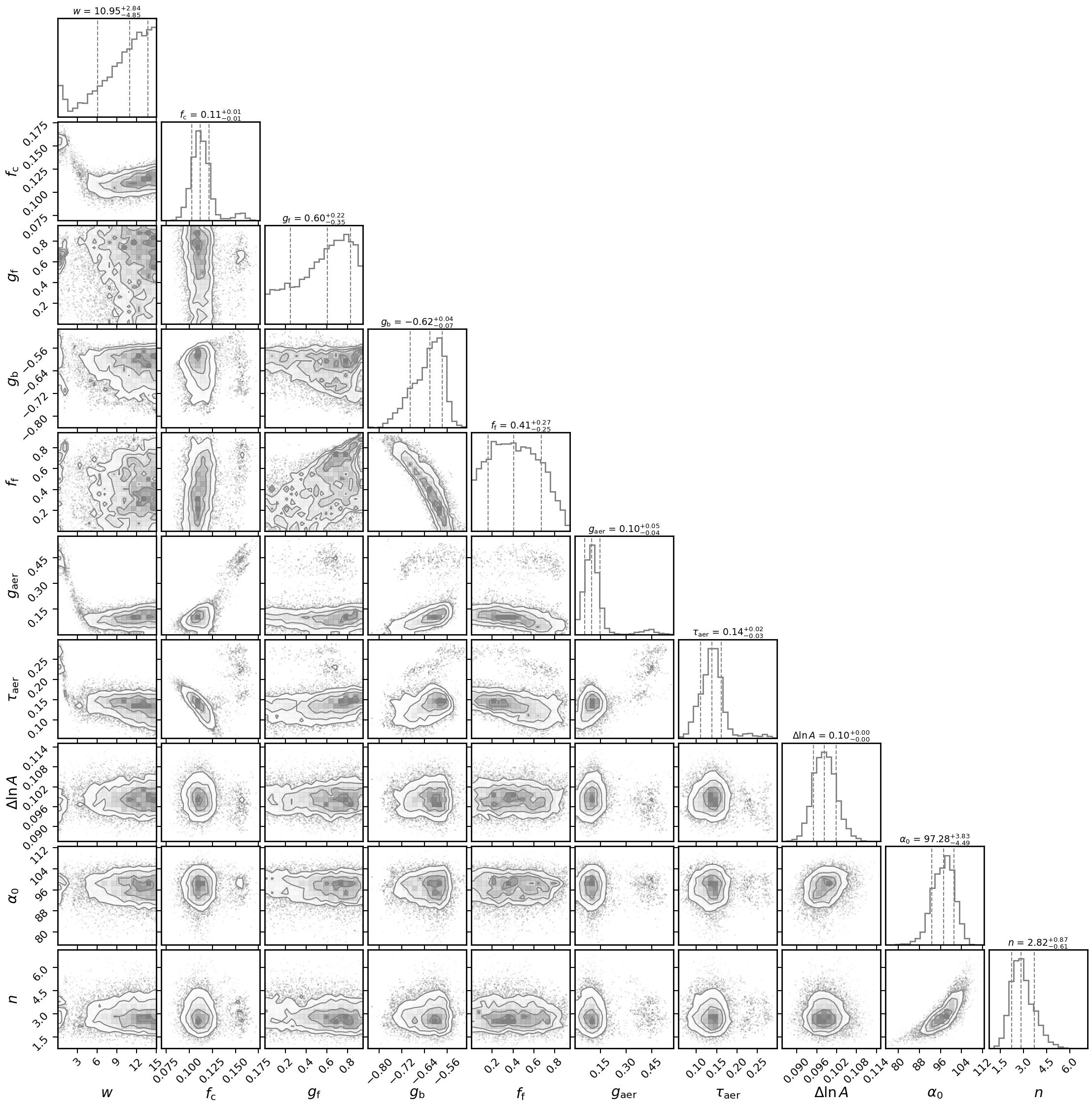}
    \caption{Corner plot for Model 05.}
    \label{fig:corner05}
\end{figure}

\begin{figure}
    \centering
    \includegraphics[scale=0.25,trim=0mm 0mm 0mm 0mm]{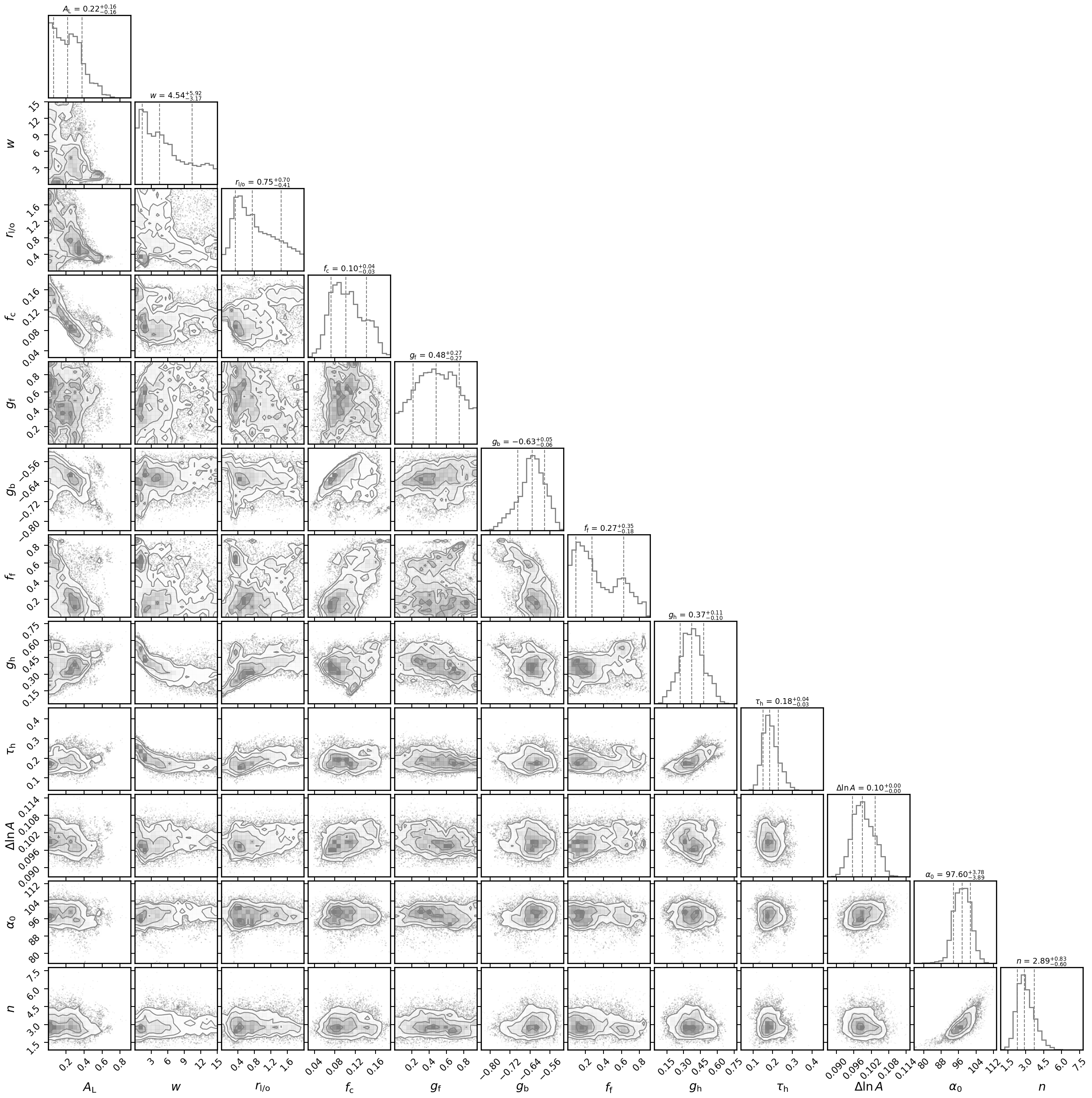}
    \caption{Corner plot for Model 07.}
    \label{fig:corner07}
\end{figure}

Corner plots for fitted parameters in Models 02, 05, and 07 (see Table~\ref{tab:models}) are provided in Figures~\ref{fig:corner02}, \ref{fig:corner05},  and \ref{fig:corner07}, respectively.

%
%%%

\clearpage

%%%
%
%\bibliography{biblist}{}
%\bibliographystyle{aasjournal}
%
%%%

%%%
%

%
%%%

\end{document}